\documentclass[11pt]{article}

\usepackage{amsmath,amssymb}
\usepackage{slashed}
\usepackage{xcolor} 
\usepackage{graphicx}
\usepackage{url}

\newcommand{\be}{\begin{equation}}
\newcommand{\ee}{\end{equation}}
\newcommand{\bea}{\begin{eqnarray}}
\newcommand{\eea}{\end{eqnarray}}
\def\Re{{\cal R \mskip-4mu \lower.1ex \hbox{\it e}\,}}
\def\Im{{\cal I \mskip-5mu \lower.1ex \hbox{\it m}\,}}

\def\tev{\,{\ifmmode\mathrm {TeV}\else TeV\fi}}
\def\gev{\,{\ifmmode\mathrm {GeV}\else GeV\fi}}
\def\mev{\,{\ifmmode\mathrm {MeV}\else MeV\fi}}

\begin{document}

\vspace{4cm}
\begin{center}
{\Large\bf{Double Higgs Production with a Jet Substructure Analysis to
  Probe Extra Dimensions
}}\\

\vspace{1cm}
 {\bf Seyed Mohsen Etesami\footnote{Corresponding author, email address: sm.etesami@ipm.ir} and Mojtaba Mohammadi
   Najafabadi}  \\
\vspace{0.5cm}
{\sl  School of Particles and Accelerators, \\
Institute for Research in Fundamental Sciences (IPM) \\
P.O. Box 19395-5531, Tehran, Iran}\\

\end{center}
\vspace{2cm}
\begin{abstract}
In this paper, we perform a comprehensive study to probe the effects
of large extra dimensions through double Higgs production in
proton-proton collisions at the center-of-mass energies of 14, 33, 100
TeV.  We concentrate on the channel in which both Higgs bosons decay
into $b\bar{b}$ pair and take into account the main background
contributions through realistic Monte-Carlo simulations. In order to
achieve an efficient event reconstruction and a good background
rejection, jet substructure techniques are used to efficiently capture
the boosted Higgs bosons in the final state. The expected
limits on the model parameters are obtained based on the invariant
mass and the angular properties of the final state
objects. Depending on the number of extra dimensions, bounds up to 
6.1, 12.5, 28.1 TeV are set on the model parameter at proton-proton collisions
with the center-of-mass energies of 14, 33, and 100 TeV, respectively. 
\end{abstract}
\vspace{1cm}

\newpage

\section{Introduction\label{sec:introduction}}

The SM-like Higgs boson discovered by the CMS and ATLAS \cite{atlashigs,cmshigs}
experiments at the LHC indicates a strong evidence for the proposed
mechanism of the spontaneous electroweak symmetry breaking (EWSB) in the
Standard Model (SM). However, further efforts are ongoing to test the characteristics
of this newly observed particle against the SM Higgs \cite{higprop} in terms
of its properties and  couplings to the SM particles. Another
important question is the EWSB behaves like what is predicted
by the SM whether or no.
To answer this question the Higgs potential has to be examined up
to higher order through measurements of Higgs boson self-couplings which
has many interesting phenomenological implications \cite{selfhig1,selfhig2,selfhig3,selfhig4,selfhig5}.
One way to explore the Higgs self-couplings is via measurements of
the di-Higgs and triple-Higgs productions at the LHC and future planned
hadron colliders \cite{spanowfsky}.

On the other hand multiple Higgs boson production also could be
used to search for new physics especially in the high invariant mass
region either for resonant or non-resonant possible physics beyond
the SM. In the resonant case, there are several interesting hypotheses
which permit new resonance decaying to Higgs pair such as Randall-Sundrum
radion \cite{rsradion} and CP-even heavy Higgs in the next to minimal
supersymmetric standard model(NMSSM) \cite{nmssm}. 
In addition to that, there are already several studies to search for new physics
through di-Higgs final state which can be found in \cite{hhh1,hhh2,hhh3,hhh4,hhh5,hhh6,hhh7,hhh8}.

In the non-resonant searches, a possibility is to 
use the di-Higgs final state at the LHC and Future Circular Colliders
(FCC) to search for large extra dimensions in Arkani-Hamed,
Dimopoulos, Dvali  (ADD) scenario \cite{ADD1}. 
They proposed the large extra dimension
scenario as a solution for the hierarchy between the scale of
electroweak and Planck scale \cite{ADD1,ADD2,ADD3}.
According to their model,  the SM fields are confined
to the 3+1 space-time dimensions while gravity can freely 
propagate into the multi-dimensional space-time $4+n_{ED}$,
where $n_{ED}$ is the possible number of extra dimensions. 
This leads to propagation of gravity field flux into
the entire $4+n_{ED}$ dimensions which leads to dilution of the power
of gravity in the common 3+1 dimensions. The reduction of the gravitational flux
can be quantified by applying the Gauss\textquoteright s law. The
result expresses the relation between the ordinary fundamental Planck
scale $M_{Pl}$ in 3+1 common dimensions and Planck scale in $4+n_{ED}$
dimensions denoted by $M_{D}$ according to the following relation:
\begin{equation}
M_{D}^{n_{ED}+2}\sim\frac{M_{Pl}^{2}}{R^{n_{ED}}}\label{eq:md}
\end{equation}
where $R$ is the size of extra dimensions. According to the ADD model motivation,
if one assumes the $M_{D}\sim M_{EW}\sim1$ TeV the size of extra dimensions
for $n_{ED}=2$ to 7 can be varied from few centimeters down to
few femto-meters.

Based on the ADD scenario,  many phenomenological 
studies have been preformed to find the possible observations in the particle colliders
\cite{grw,hlz,hewett}. 
In these works, graviton in the multi-dimensional
representation equivalently interpreted as towers of massive Kaluza-Klein
(KK) modes or $G_{kk}$which can couple to the SM
particles through SM energy-momentum tensor. The resulting effective
model provides different experimental signatures such as virtual exchange of graviton 
and direct graviton emission at colliders.

Although the coupling of each KK mode $G_{kk}$ with the SM gauge bosons or
fermions is suppressed by the Planck scale, the summation over all
KK-modes with tiny mass splitting $\Delta m_{KK} \sim 1/R$ compensates
for the suppression of Planck scale. It is notable that such a mass
scale is not observable by the current experiments due to very limited
resolutions. In order to avoid of any
divergency in the production cross sections arising from summation
over the infinite number of KK modes, an ultraviolet cutoff scale
$\Lambda_{UV}$ is essential to regulate the processes.
The extra dimension model is a low energy effective theory which is valid
below the onset of quantum gravity scale, denoted by the scale $M_{S}$.
Throughout this analysis,  the cutoff scale of the effective theory
$\Lambda_{UV}$ is chosen to be equal to $M_{S}$.
In general, $M_{S}$ is different from the
Planck scale in the presence of extra dimensions $M_{D}$ but it is
related to $M_{D}$ according to the following relation \cite{rr,lhc5}:
\begin{eqnarray}
M_{S} = 2\sqrt{\pi}\left[\Gamma(\frac{n_{ED}}{2}) \right]^{1/(n_{ED}+2)}\times M_{D},
\end{eqnarray}

The cross sections of processes in the ADD model are usually
parametrized using the parameter $\eta_{G}$ which is equal to
$F/M_{S}^{4}$ where $F$ is a dimensionless parameter which takes
different forms in different conventions including the GRW \cite{grw}, HLZ \cite{hlz}
and Hewett \cite{hewett} conventions.
In the HLZ conventions, $F$ is expressed as a function
of $M_{S}$ and number of extra dimensions:
\begin{eqnarray}
F =
\left\{
	\begin{array}{ll}
		\log(\frac{M_{S}^{2}}{\hat{s}})  & \mbox{if } n_{ED} =
                2 \\
		\frac{2}{n_{ED}-2} & \mbox{if } n_{ED} > 2,
	\end{array}
\right.
\end{eqnarray}
where $\sqrt{\hat{s}}$ the center-of-mass energy of the hard process
which is approximately equal to the di-Higgs invariant mass in this analysis. 
In the GRW convention, $F$ is equal to one and the scattering
amplitude of graviton mediated processes can be
parametrized in terms of a single parameter $\varLambda_{T}$ \cite{grw}:
\begin{eqnarray}
\mathcal{A} =
\frac{4\pi}{\Lambda_{T}^{4}}\mathcal{T}~,\text{for}~n_{ED} > 2,
\end{eqnarray}
where $\mathcal{T}$ is a function of energy-momentum tensor. The GRW
and HLZ conventions can be related using $M_{S}^{4} = F\times
\Lambda_{T}^{4}$. In this work, the results are presented in both
conventions.
\\
There are different possibilities which can be used to study the effects of 
large extra dimensions. Some experimental tests of the ADD model are
mentioned here. \\
\textit{Gravitational law:} the Newton's gravitational force will be
modified in the ADD model framework at distances shorter than the size
of the extra dimensions. At $95\%$ CL, the size of the extra dimension
above $37\mu$m has been excluded. This is corresponding to an
exclusion of $M_{D}$ below 1.4 TeV for two extra dimensions \cite{n1,n2}. \\
\textit{Collider experiments:} as mentioned before, large extra dimension leads to direct production of
gravitons at particles colliders as well as enhancements in the cross
sections of some SM processes due to virtual graviton exchange.  
Experimental limits on the extra dimensions have been set by different
experiments including the HERA \cite{hera1,hera2}, LEP \cite{lep1,lep2,lep3,lep4}
and Tevatron \cite{tevatron1,tevatron2}. At the LHC, the ADD model has
been probed in diphoton, dilepton, monophoton, and monojet channels in both CMS
and ATLAS experiments \cite{lhc1,lhc2,lhc3,lhc4,lhc5,lhc6,lhc7,lhc8,lhc9}.
The most stringent collider limits on  $\varLambda_{T}$ come from the LHC run at the
center-of-mass energy of 8 TeV from dilepton and monojet events which
are 4.0 TeV and 3.74 TeV, respectively \cite{lhc7,lhc8,lhc9} .
Another interesting signature of the large extra dimensions at
collider experiments is the black hole production
\cite{bh1,bh2}. \\
\textit{Cosmological and astrophysical constraints:}
cosmological and astrophysical observations provide strong bounds on
the large extra dimension model parameters. Star cooling, $\gamma$ ray
diffusion and universe expansion during the big bang nucleosynthesis
are examples of astrophysical and cosmological implications by which
the ADD model are constrained. More details can be found in
\cite{aa1,aa2,aa3,aa4}.\\
There are other studies on the consequences of the large extra dimensions
in the electroweak precision test, neutrino physics etc in the
literature \cite{e1,e2,e3,e4,e5,e6}.

So far, the theoretical cross section of di-Higgs production in the
context of large extra dimension has been calculated in the
$\gamma\gamma$, $e^{-}e^{+}$, and $pp$ colliders \cite{hh1,hh2,hh3,hh4}.  
In this work, we perform a detailed search for the ADD model based on the
di-Higgs production in proton-proton collisions at the LHC and FCC at the center-of-mass energies of 14, 33, 100
TeV. All major backgrounds are taken into account and the effects of a
CMS-like detector is considered.  The jet substructure technique is utilized to
capture boosted Higgs bosons and to reach a reasonable background
rejection and efficient event reconstruction.

\section{Double Higgs production in $pp$ collisions}

\begin{figure}
\begin{centering}
\includegraphics{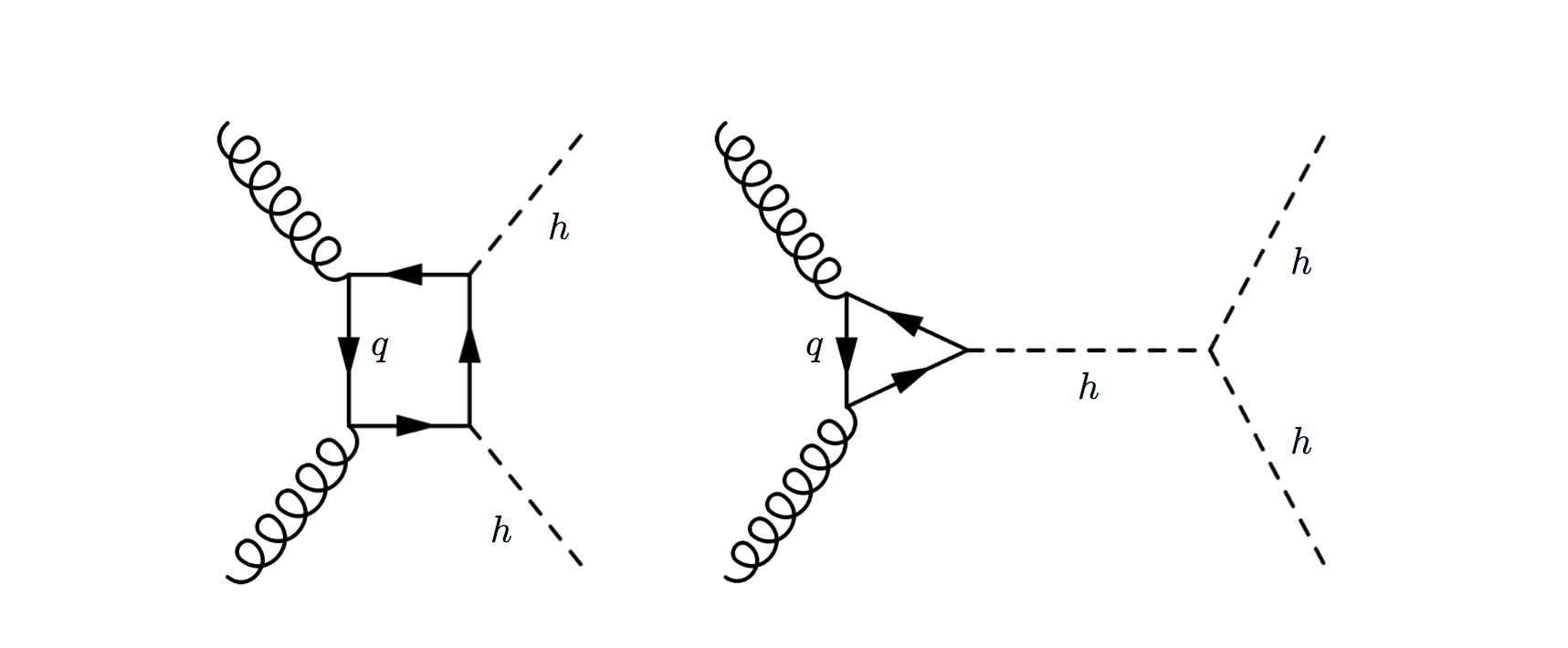}\protect\caption{SM loop diagrams for Higgs pair production
via gluon fusion.\label{fig:smdiagrams}}
\par\end{centering}
\end{figure}

The double Higgs boson production at hadron colliders within the SM has been studied in \cite{sm1,sm2}.
The representative Feynman diagrams for production of two Higgs bosons
at hadron colliders are presented in Fig.\ref{fig:smdiagrams}. 
The di-Higgs final state proceeds through $gg$ fusion via quark loop
diagrams and $q\bar{q}$ annihilation.
The main contribution to the total production rate comes from
the loop diagram involving mostly the top quark in the  $gg$ fusion.
Due to larger parton distribution functions of the gluon and very
small Yukawa couplings of the Higgs boson with light quarks,  the
dominant contribution of the di-Higgs production
comes from gluon-gluon fusion involving the triangle and box diagrams.
The total cross section of di-Higgs at next-to-leading (NLO) order calculated assuming the top quark mass $M_{t}=$173.1 GeV,
bottom quark mass $M_{b}=5$ GeV, Higgs boson mass $M_{H}=125$ GeV, $\alpha_{s}^{LO}(M_{Z}^{2})=0.13939$
and $\alpha_{s}^{NLO}(M_{Z}^{2})=0.12018$ at $\sqrt{s} =$ 14 TeV, 33 TeV and 100
TeV  are 33.89 fb, 207.29 fb and 1417.83 fb, respectively
\cite{dihigxsec}. For cross section calculation,  the CTEQ66
\cite{cteq} PDF set is used.
 An interesting aspect of di-Higgs production is the destructive
 interference between the box and the triangle contributions shown in Fig.\ref{fig:smdiagrams}.
It is worth mentioning that the destructive interference is not
negligible  so that it leads to a 
reduction of around $50\%$ in the production rate \footnote{The
  amount of reduction in the total cross section due to the
  interference term depends on the center-of-mass energy of the
  collision. At $\sqrt{s} = 14$ TeV, it amounts to $~50\%$. }.

\begin{figure}
\begin{centering}
\includegraphics{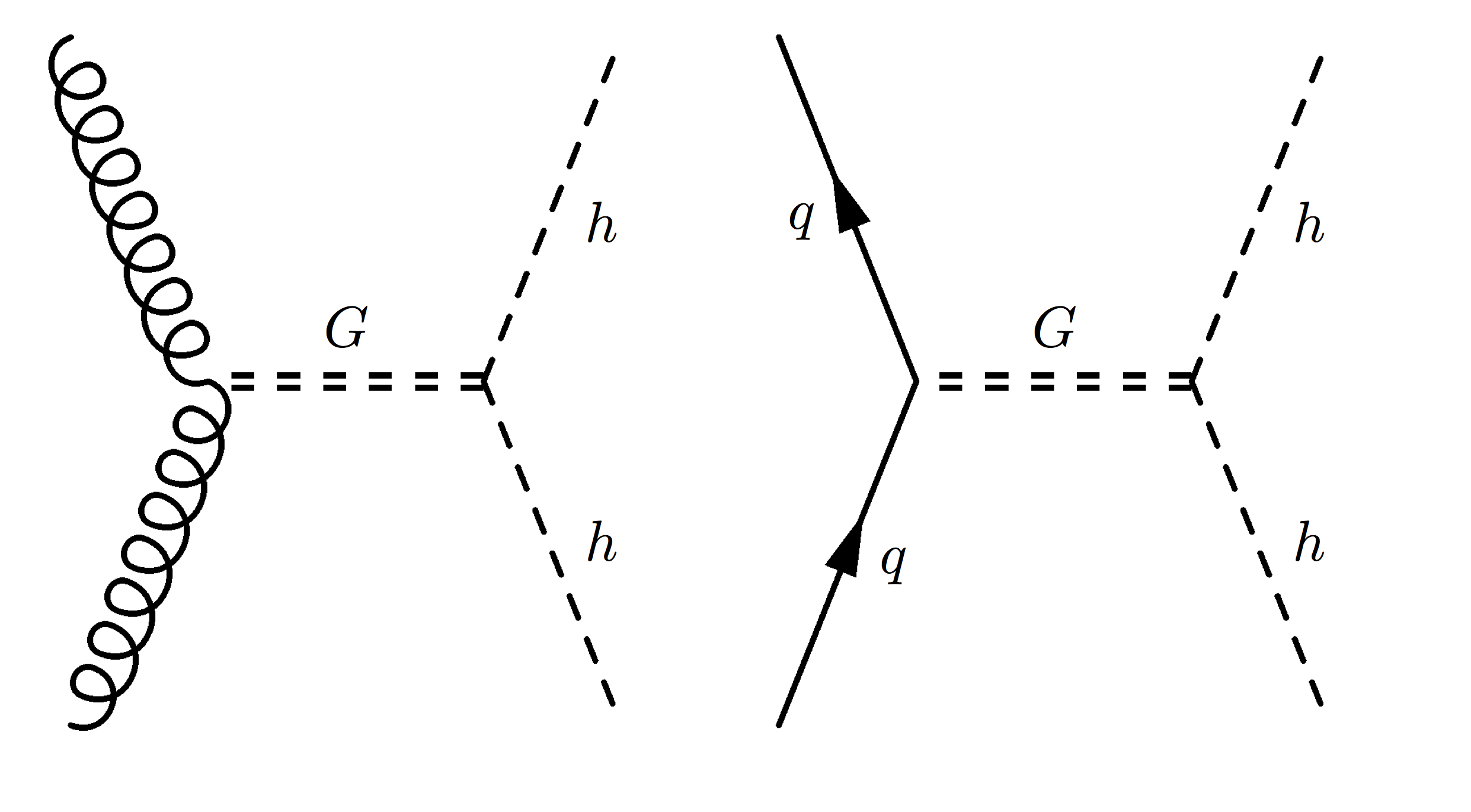}\protect\caption{Tree level di-Higgs
  production diagrams with virtual  graviton exchange in the ADD model.\label{fig:adddiag}}
\par\end{centering}
\end{figure}

Within the ADD model, the double Higgs production occurs at tree level through
both $gg$ fusion and $q\bar{q}$ annihilation via $s$-channel. The
representative Feynman diagrams are presented in Fig.\ref{fig:adddiag}.
As it can be seen, in the ADD model two Higgs bosons are produced at tree level
via virtual gravitons exchanges.  The presence of the new diagrams lead to
increase the total rate of Higgs pair with respect to the SM rate.
On the other hand, due to mediating spin-2 graviton one
expects different kinematical properties between the Higgs
pair from the SM and ADD model. These issues will
be discussed more in the next sections. 

\begin{figure}
\begin{centering}
\includegraphics[scale=0.6]{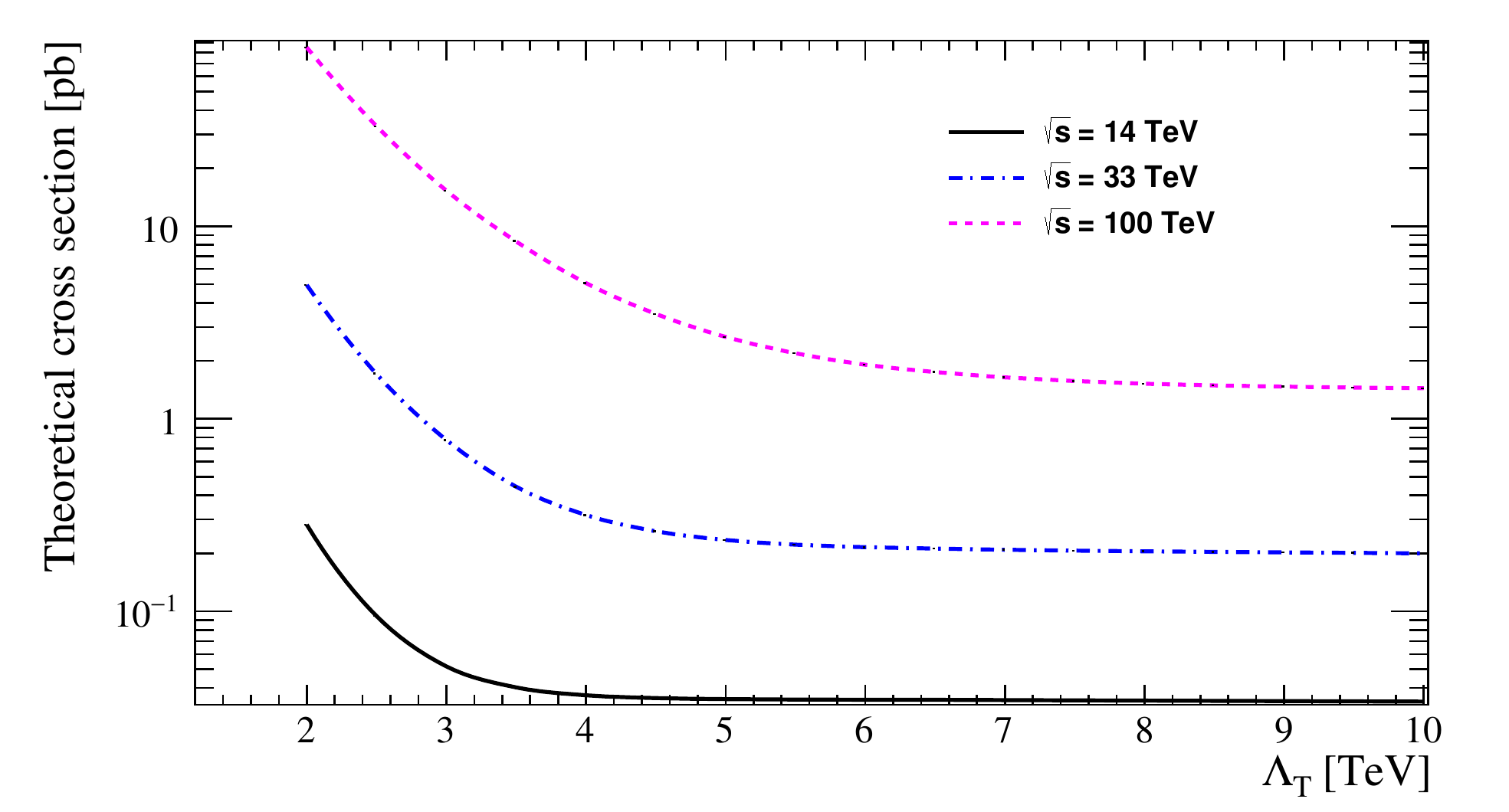}
\par\end{centering}
\centering{}\protect\caption{Cross section of pair Higgs production at three
different center-of-mass energies of 14, 33 and 100 TeV as a function of
$\Lambda_{T}$ in the GRW convention. It can be seen that cross sections
tend to the SM ones as $\Lambda_{T} \rightarrow \infty$. \label{fig:thxsec}}
\end{figure}

In this work, Sherpa \cite{sherpa} event generator is used to generate
the di-Higgs events and to calculate the cross sections in the ADD model. Figure \ref{fig:thxsec}
shows the calculated cross section of Higgs boson pair production at three different
center-of-mass energies of 14, 33 and
100 TeV as a function of ADD model parameter $\Lambda_{T}$ in the GRW
convention.  As expected, the total production cross section of two
Higgs bosons grows significantly with respect to the expectation of the
SM at the three center-of-mass energies. Due to
larger phase space and PDFs, the cross section increases with
increasing the center-of-mass energy. Because the $gg$ and
$q\bar{q}$ interactions with gravitons are suppressed by the Planck scale in
$4+n_{ED}$ dimensions, the cross section is expected to decrease with
increasing the ADD model scale $\Lambda_{T}$ hence the cross section
goes to the SM value when $\Lambda_{T} \rightarrow \infty$.

\begin{figure}
\centering{}
\includegraphics[scale=0.35]{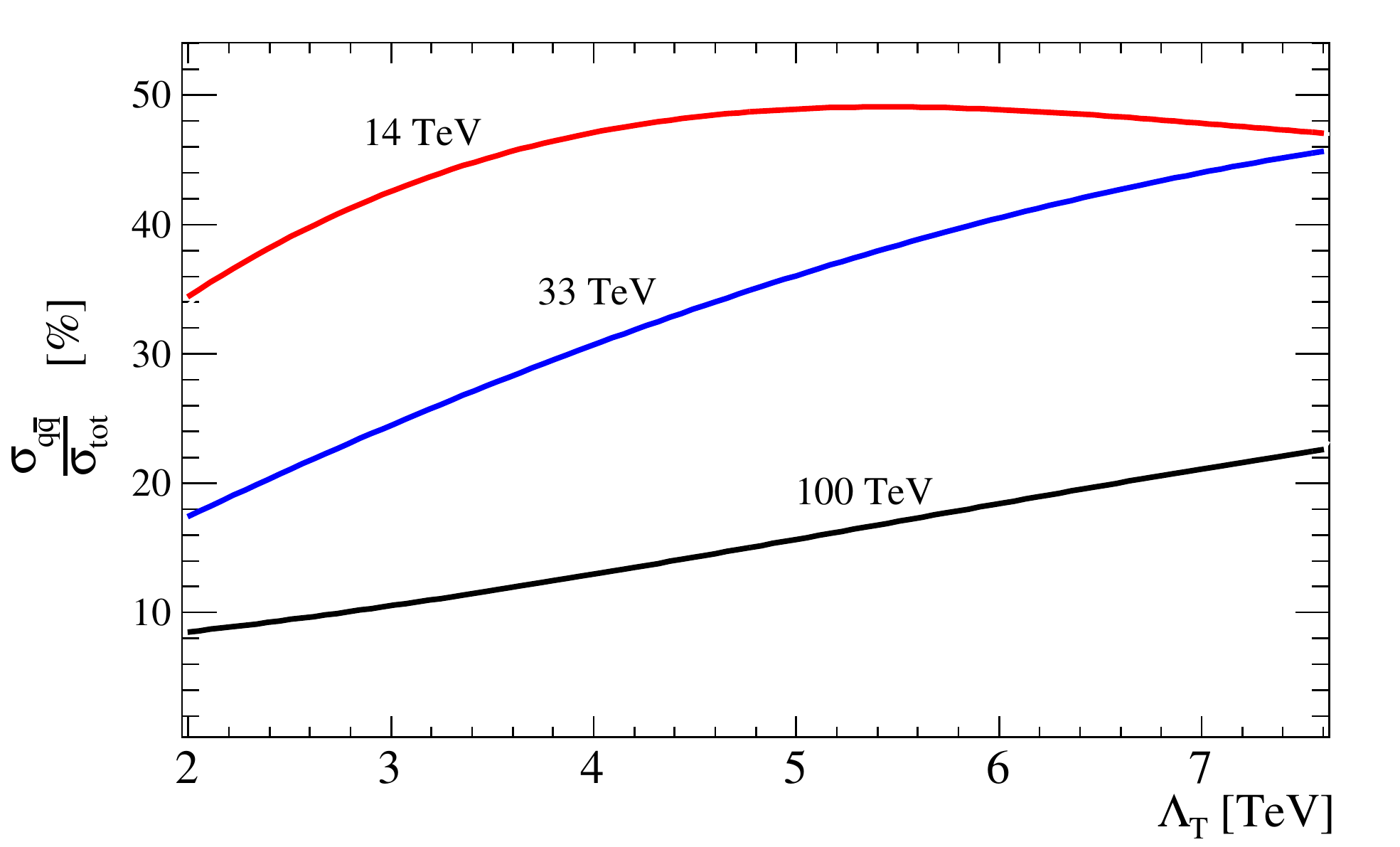}
\includegraphics[scale=0.35]{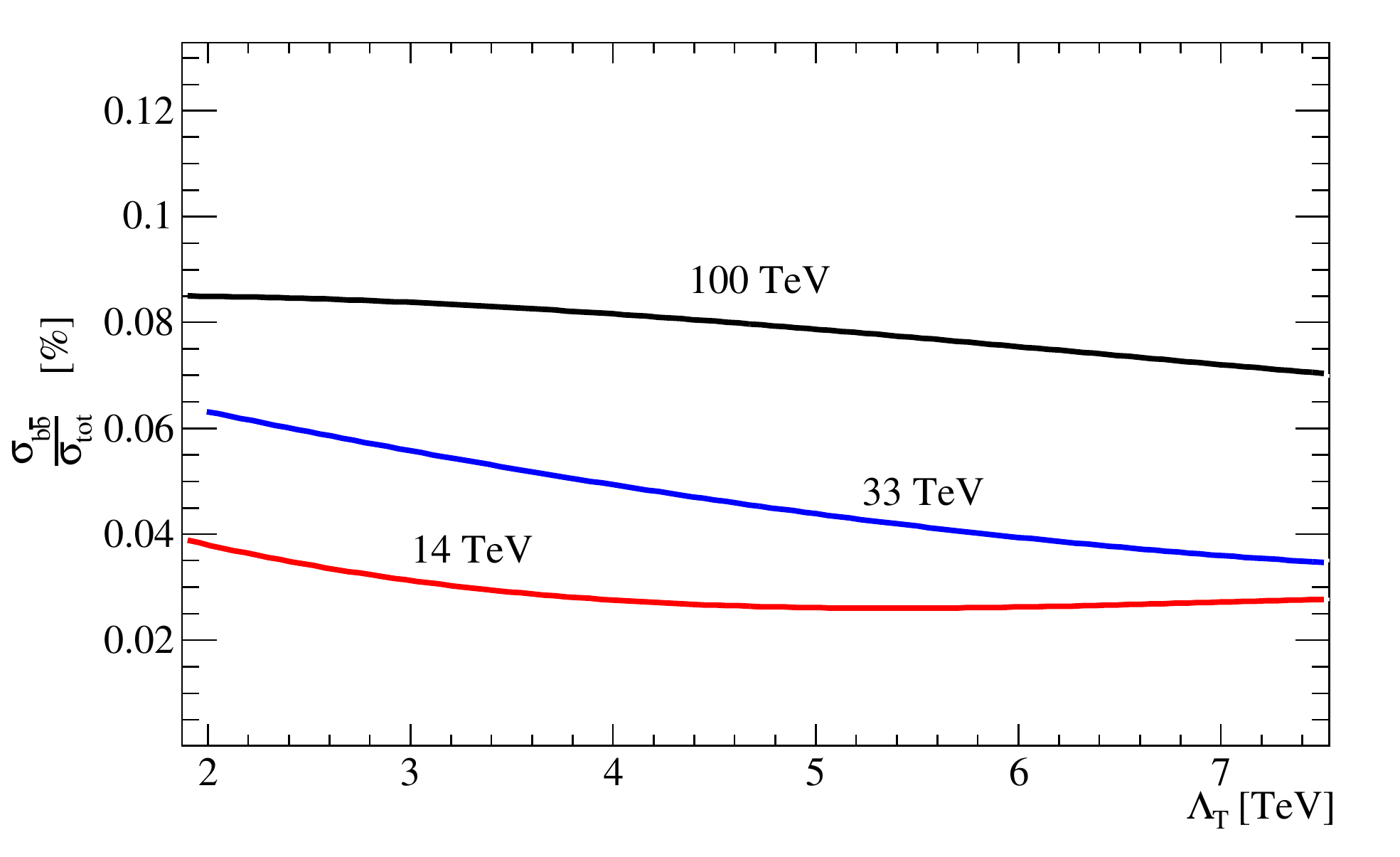}
\protect\caption{Ratio of partonic cross section over the total cross section for the
quark anti-quark initial state.\label{fig:xsecqq}}
\end{figure}

\begin{figure}
\begin{centering}
\includegraphics[scale=0.5]{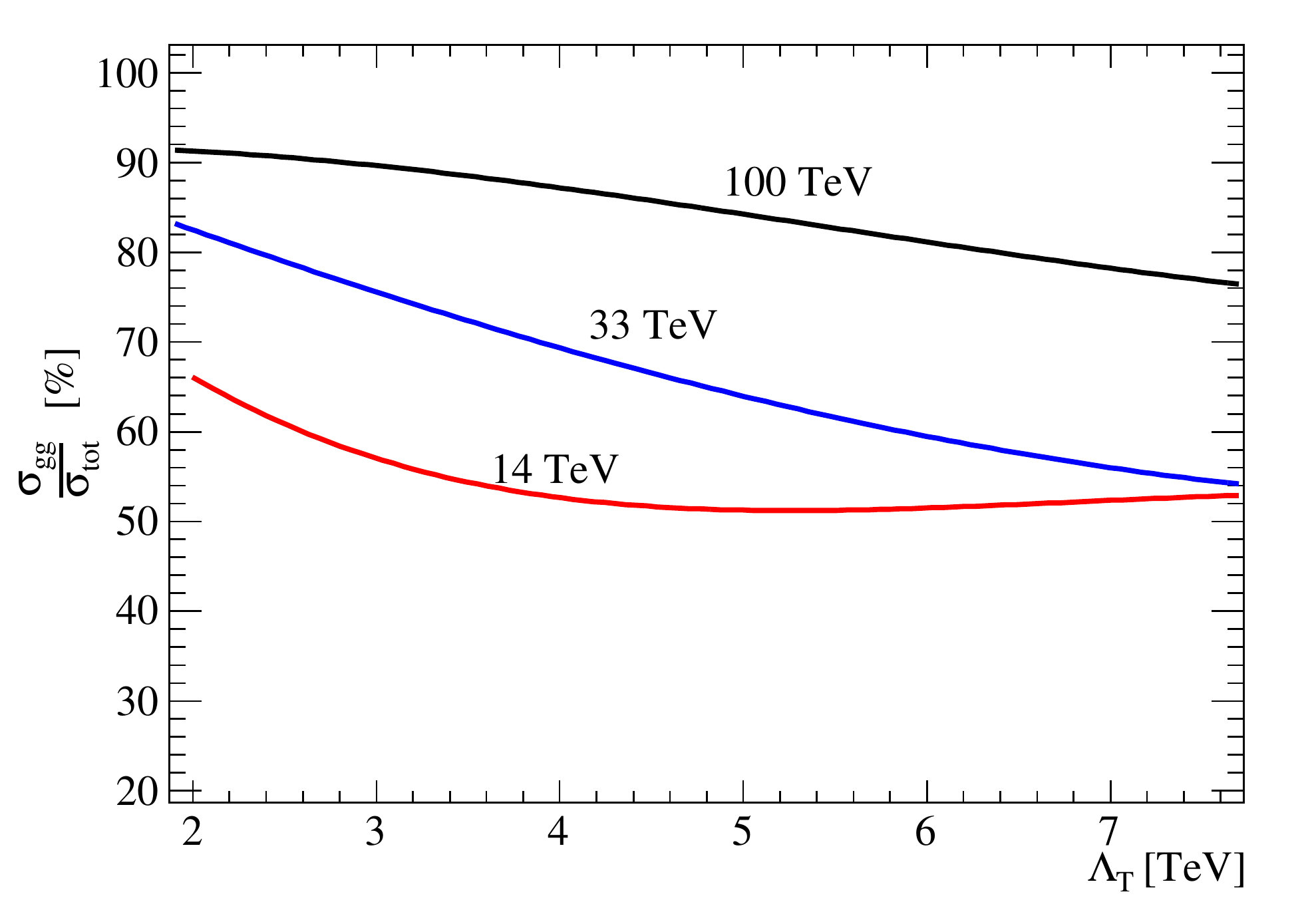}\protect\caption{Ratio of partonic cross section over the total cross section for the
gluon-gluon initial state.\label{fig:xsecgg}}
\par\end{centering}
\end{figure}

Figure \ref{fig:xsecqq} shows the ratio of di-Higgs cross section from
$q\bar{q}$ annihilation and only $b\bar{b}$ to the total cross section
versus the ADD model parameter $\Lambda_{T}$ at three collision energies of 14, 33, and 100 TeV.
The contribution of the $gg$ fusion is shown in 
Fig.\ref{fig:xsecgg}. As expected the main contribution is coming from
the $gg$ fusion with the amount of more than $60\%$ and $90\%$ at
$\Lambda_{T} = 2$ TeV at the
center-of-mass energies of 14 and 100 TeV, respectively. With
increasing the  center-of-mass energy, the contribution from $gg$ fusion is
increased. It is interesting to note that again with increasing the
center-of-mass energy, the b-quark parton distribution function is
increased which leads to larger contribution from $b\bar{b}$
annihilation at larger energies.

We close this section by mentioning that the ADD model leads to
produce considerable number of Higgs boson pairs in $pp$
collisions. We will see that the increment in number of Higgs pairs in
particular occurs at the large invariant mass of the two Higgs system
$M_{HH}$. Such an effect will be used as a tool to search for the ADD
model and constraining the model parameters at different collision energies.
The details of the analysis are described in the next sections.


\section{Analysis Method}

In this section,  we  explain the analysis procedure that is followed for generating and analyzing
the signal of ADD model in $pp$ collisions at the LHC with
$\sqrt{s}=14$ TeV and FCC.  Throughout this
analysis, we use a CMS-like experiment characteristics for simulating
the effects of detector and similar statistical tools for obtaining the
exclusion limits as the CMS experiment.  We only focus on the Higgs bosons decay into
$b\bar{b}$ pairs. This leads to have a final state containing four
jets originating from the hadronization of b-quraks.

\subsection{Event generation}

The ADD signal events are generated using Sherpa version $2\_1\_1$
\cite{sherpa} in the GRW convention. Sherpa also performs parton showering and
hadronization processes. The background with most similarity to the signal which can be interpreted
as the irreducible background comes from the SM di-Higgs explained previously.
The SM di-Higgs event generation is done with the MadGraph 5 \cite{madgraph5,hhsm}
and Pythia \cite{pythia6} is used to perform parton showering and
hadronization.  
The remaining main background processes are QCD multijets, $Zb\bar{b}$
$ZZ$, $ZH$, $t\bar{t}$, single top,  $W^{+}W^{-}$, $W+jets$ which are generated
using Sherpa including the showering and
hadronization. 
The contribution of multijet QCD background is difficult to reliably estimate due to
large production rate.  In reality, the determination of the QCD
multijet contribution requires computing resources and/or employing a data-driven
technique  which is beyond the scope of the current work. 
We generate several QCD multijet samples in various
bins of invariant mass of the final state partons with large amount of events in each
bin. 

A full and real detector effects simulation must be performed by the
experimental collaboration however we use Delphes \cite{delphes} as 
the tool to estimate the response of the detector. It considers a
modeling of the CMS detector performances as explained in \cite{cms}.
In this study, the effects of pileup and underlying events are not
taken into account.

Finally, we should mention that the signal samples are generated
with Sherpa in the GRW convention
for the various values of the ADD model parameter $\Lambda_{T}$ and
all simulated samples are generated for the three possible scenarios
of the center-of-mass energies of proton-proton colliders,
14 TeV, 33 TeV and 100 TeV.

\subsection{Analysis details}

We perform the analysis of the simulated events on the stable final
state particles. The selection of the events 
is designed in such a way to find the $HH$
events with subsequent decay of $HH\rightarrow b\bar{b}b\bar{b}$.

\begin{figure}
\begin{centering}
\includegraphics[scale=0.5]{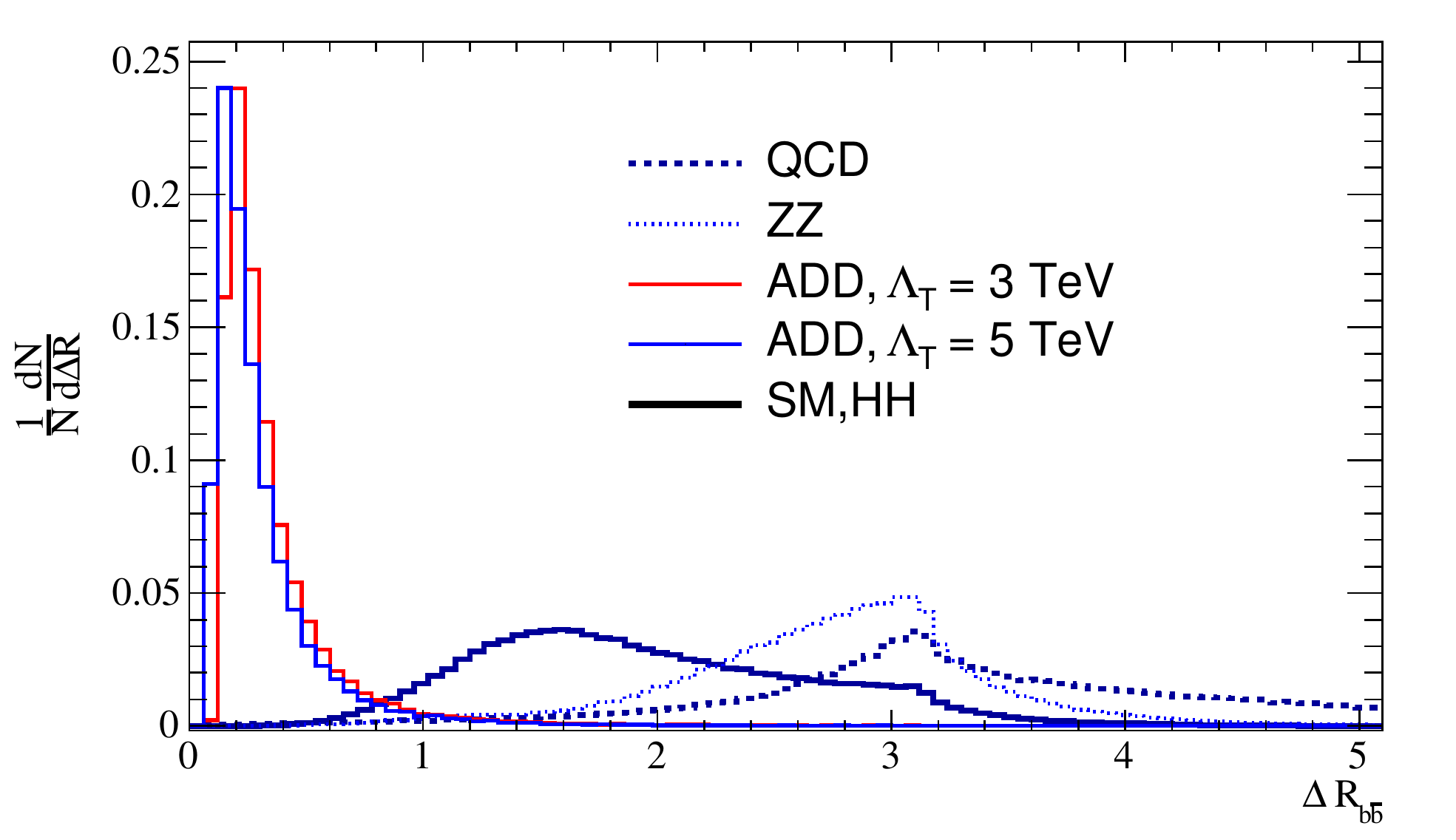}
\end{centering}
\centering{}\protect\caption{Normalized distribution of the $\Delta R$ between two bottom quarks
coming from decay of each Higgs for the two samples of signals and
the potential backgrounds.\label{fig:dltar14}}
\end{figure}

Before going further, one of the special characteristic of
the signal events, which leads to employ
particular strategies for events reconstruction and selection, is considered .
In Fig.\ref{fig:dltar14},  the normalized distribution
of  $\Delta R$ \footnote{$\Delta R$ is the angular separation of
  $b$ and $\bar{b}$ quarks in the
$\eta-\varphi$ plane which is defined as: $\Delta
R_{b\bar{b}}=\sqrt{(\eta_{b}- \eta_{\bar{b}})^{2}+(\phi_{b}-\phi_{\bar{b}})^{2}}$} 
between two bottom quarks coming from the decay of
each Higgs boson for two samples of signal with $\Lambda_{T} = 3,5$
TeV and some backgrounds is shown.
As it can be seen in Fig.\ref{fig:dltar14},  signal events tend to
reside at very small values of $\Delta R$ contrary to the SM
backgrounds.

\begin{figure}
\begin{centering}
\includegraphics[width=12cm,height=6cm]{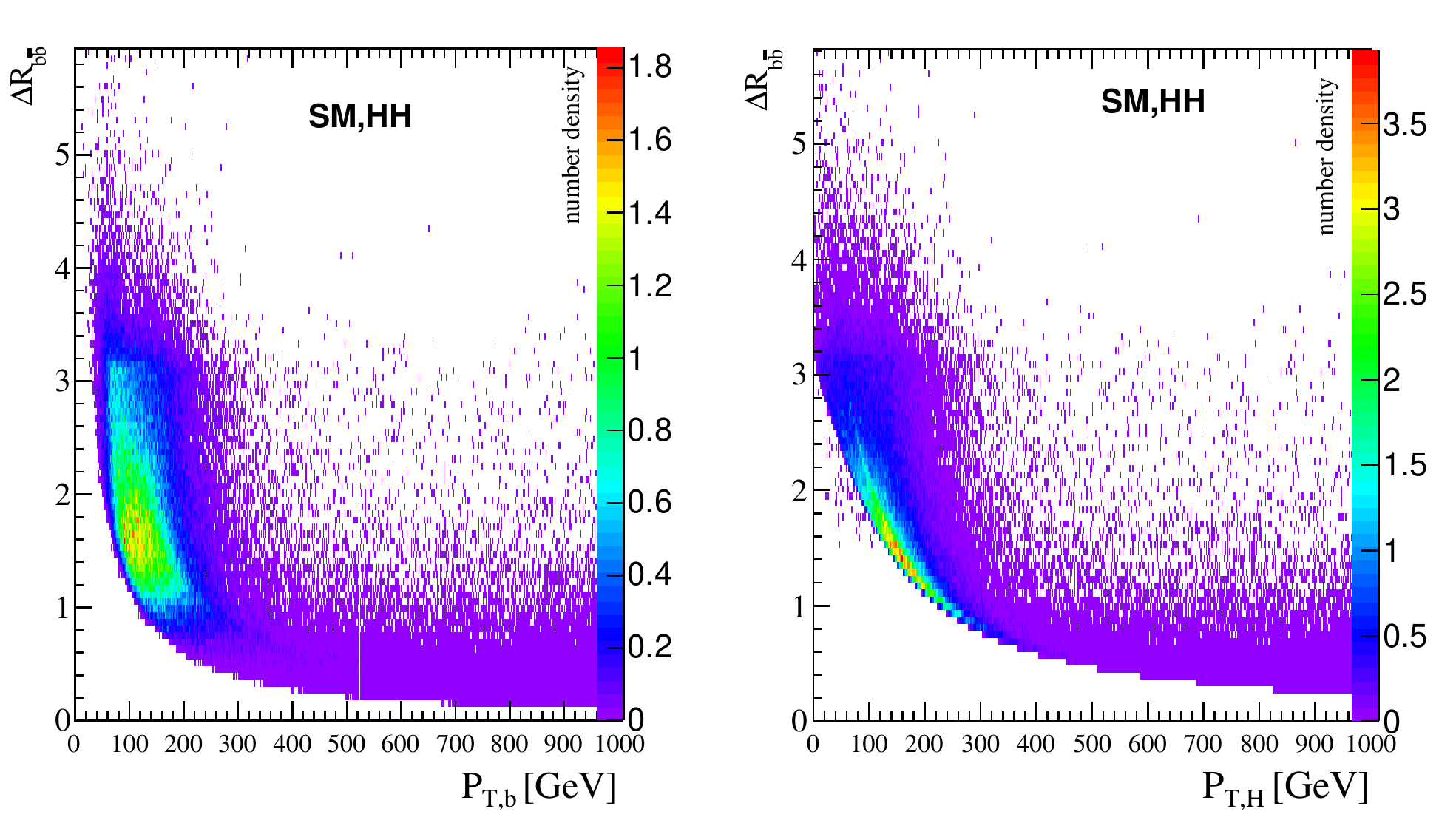}\protect\caption{The transverse
  momentum of the hardest bottom quark and the Higgs boson transverse momentum as a function of $\Delta R$
for the SM di-Higgs events. \label{fig:deltarpt}}
\par\end{centering}
\end{figure}

In the events of signal with very large di-Higgs invariant
mass, Higgs bosons are Lorentz-boosted particles which decay differently from
the topological point of view compared to the Higgs bosons which are produced almost
at rest. The angular separation of a $b\bar{b}$ pair coming from a
Higgs boson can be approximated as:
\begin{equation}
\Delta R_{b\bar{b}}\simeq\frac{1}{\sqrt{x(1-x)}}\frac{m_{H}}{P_{T}}\label{eq:deltar}
\end{equation}
where $P_{T}$ is the transverse momentum of the Higgs boson, $x$ and
$1-x$ are the momentum fractions of the $b$ and $\bar{b}$ quarks. The
larger Higgs $P_{T}$ the smaller angular separation of $b\bar{b}$ pair.
Figure \ref{fig:deltarpt} shows two dimensional plots of $\Delta R_{b\bar{b}}$
versus the Higgs boson $P_{T}$ and the hardest 
$b$ quark $P_{T}$ in a Higgs boson decay for the SM process of $pp \rightarrow HH \rightarrow b\bar{b}b\bar{b}$.
The plots of Figure \ref{fig:deltarpt} confirm that as we go to the
boosted region (events with large transverse momentum of Higgs or
large transverse momentum of b quark),
the $\Delta R_{b\bar{b}}$ decreases. It means that the boosted Higgs
bosons produce one collimated jet with substructure. This can originate
from two reasons: first is that the decaying Higgs boson has an energy
several times larger than the Higgs boson mass in the laboratory frame
and second reason is the difference between the mass of Higgs and b
quark is large.

\begin{figure}
\centering{}
\includegraphics[width=7cm,height=5cm]{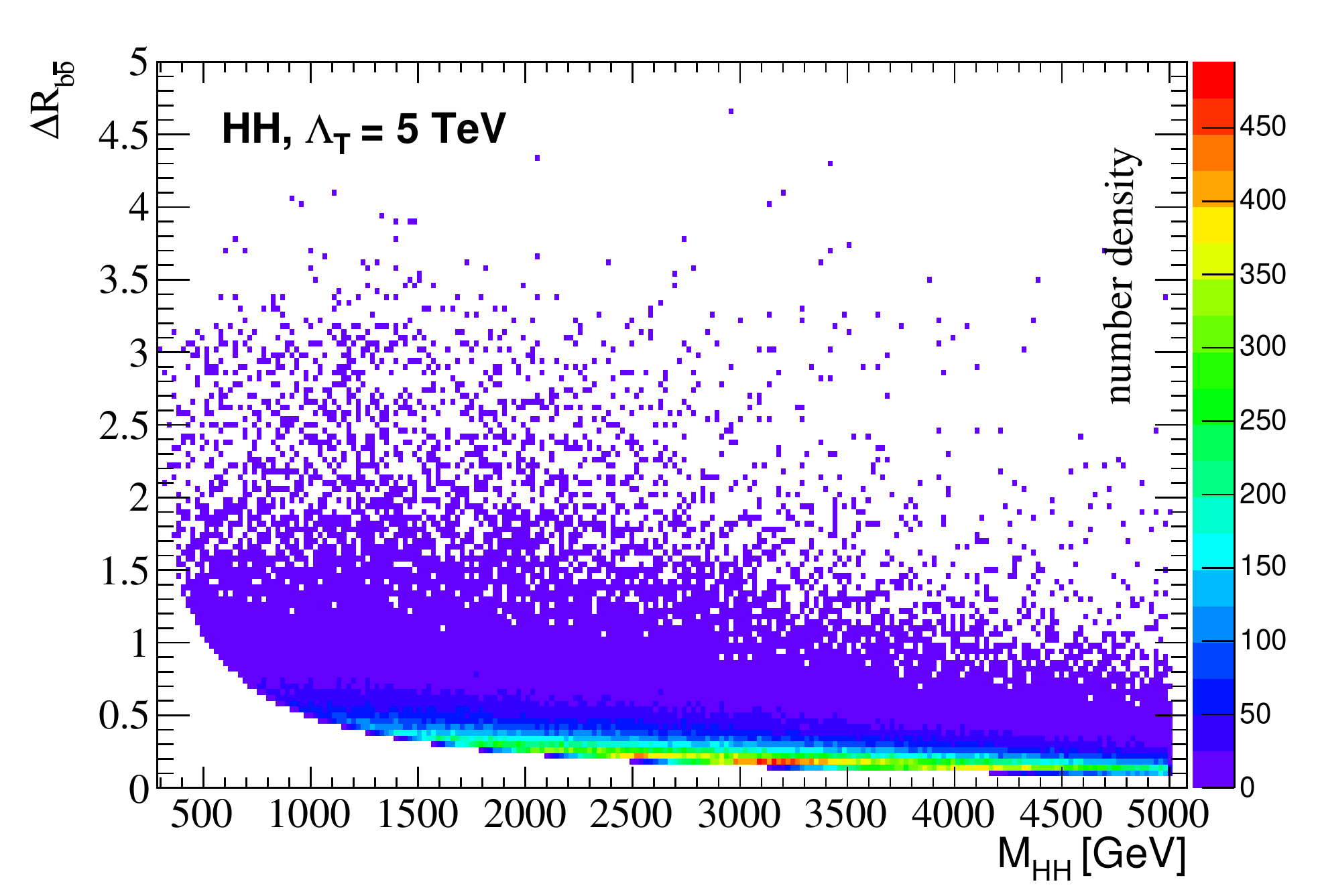}
\includegraphics[width=7cm,height=5cm]{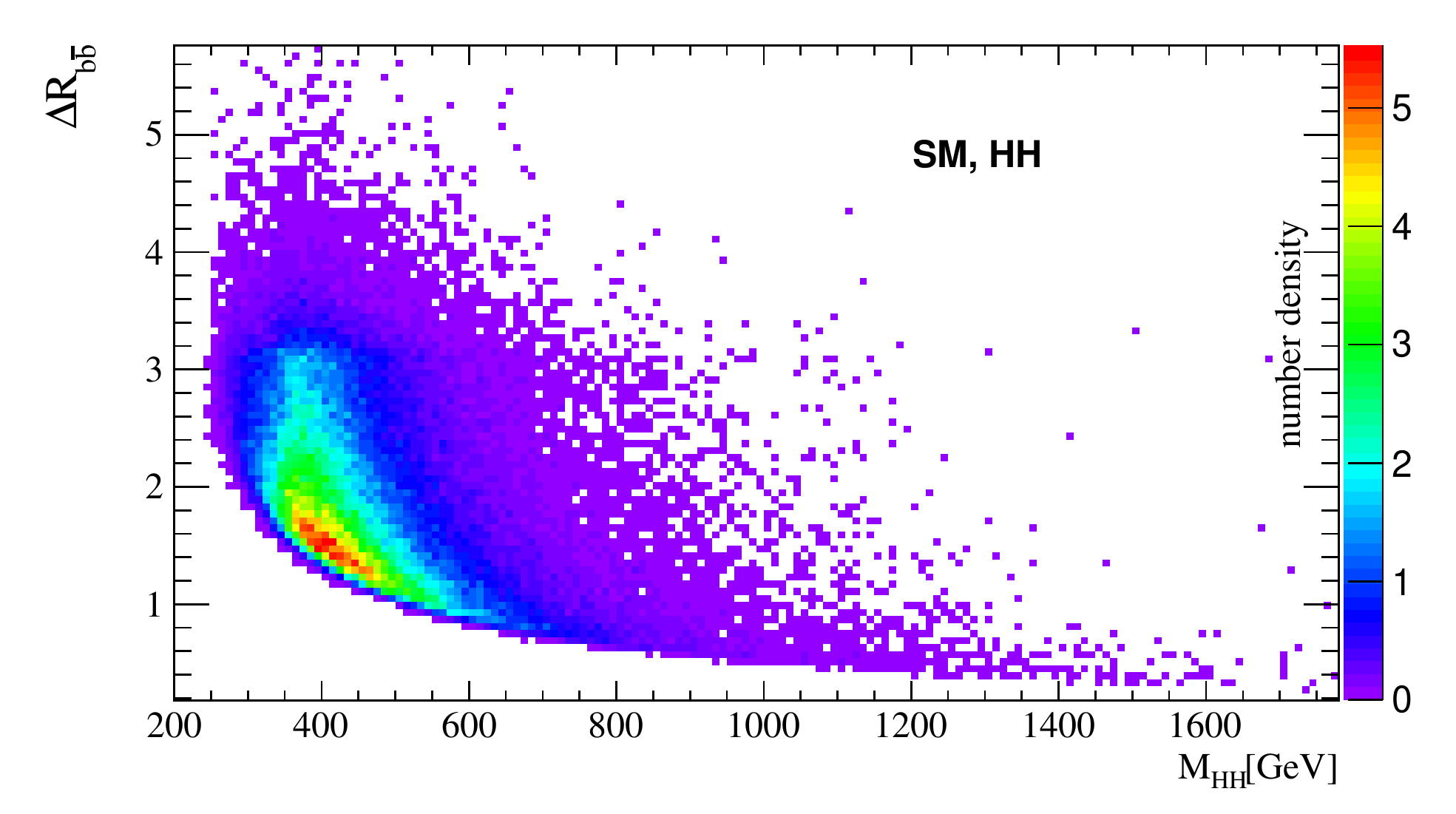}
\protect\caption{Invariant mass of di-Higgs as a function of the $\Delta R$ between
the each b-jet pairs for the signal with $\Lambda_{T}=$ 5 TeV (left)
and for the SM di-Higgs events (right).\label{fig:invdeltar}}
\end{figure}

As the di-Higgs invariant mass is an important quantity which will be
used to separate ADD signal events from the backgrounds,
$\Delta R_{b\bar{b}}$ is also presented versus the di-Higgs invariant
mass for signal and SM di-Higgs events in Figure \ref{fig:invdeltar}.
Obviously, the ADD signal events prefer to be distributed in the large
invariant mass region of the di-Higgs system which is not the case for
the SM di-Higgs events. Another observation from the Figure
\ref{fig:invdeltar} is that with increasing the di-Higgs mass of the
signal events, $\Delta R_{b\bar{b}}$ decreases and falls even below 0.4.
More accurately,  the SM di-Higgs events are distributed at di-Higgs
invariant mass smaller than 1 TeV and peak at $\Delta R=1.5$ while for
the signal, di-Higgs invariant mass peak at values greater than 3 TeV  
and around $\Delta R=0.3$.

The kinematics of the Higgs boson decay products is
categorized by two types of event topologies. The first category
consists of Higgs boson pairs which are produced near the
threshold. In this type of events (normal events),  each parton is matched to a single
jet. In the second category, Higgs bosons are
 produced with a large Lorentz boost, resulting in collimated jets that might
cluster into one jet.  These events are referred as boosted events
which need to be treated differently from the normal events.

\subsection{Boosted jet reconstruction \label{sec:Pre-Event-Selection-and}}

As it has been mentioned previously,  reconstruction of b-jets
specially for the signal events due to the existence of highly boosted Higgs
bosons are crucial.  According to Eq.\ref{eq:deltar} if the Higgs
bosons have transverse momentum larger than $P_{T}>500\: GeV$
and if approximately both b-jets carry the same fractions of the Higgs
boson momentum,  the angular separation between two
b-jets is $\Delta R<0.5$. As a consequence, common jet reconstruction 
which usually is done with the cone size of $\Delta R=0.4-0.5$ would not be applicable
for the most of  signal events. Therefore, an alternative way of fat
jet algorithm is used \cite{fatjetSalam} for these boosted events.

Now,  the jet substructure analysis is described together with
its application on our signal with four boosted b-jets in the
final state.  We reconstruct the fat jets using the Cambridge/Aachen
(CA) jet algorithm \cite{ca1,ca2} assuming special jet cone size of
$R=1.2$. Then to identify the
boosted Higgs bosons,   the  procedures described in the fat jet
reconstruction algorithm \cite{fatjetSalam} is performed as the following. First, large
radius or a fat jet $J$ is split into two sub-jets $J_{1}$ and
$J_{2}$ with masses $m_{J_{1}}>m_{J_{2}}$. Then a significant mass drop
  of $m_{J_{1}}<\mu_{MD} m_{J}$ with $\mu=0.667$ is required. 
$\mu_{MD}$ is an arbitrary value that shows the degree of the mass
drop.  In order to avoid the inclusion of high $P_{T}$ light jets,
two sub-jets have to be symmetrically split by satisfying:
\begin{equation}
\frac{min(P_{T,\, J_{1}}^{2},P_{T,\, J_{2}}^{2})}{m_{J}^{2}}\Delta R_{J_{1},\, J_{2}}^{2}<y_{cut}
\end{equation}
where $P_{T,\, J_{1}}^{2}$and $P_{T,\, J_{2}}^{2}$ are the square
of the transverse momentum of each sub-jet and $y_{cut}$ is one
of the parameters of the algorithms which determines the limit of asymmetry
between two sub-jets. 
Finally, if the criteria in the above steps are not fulfilled,  we
take $J=J_{1}$ and return to the first step to perform decomposition. 
All the above steps are followed by a filtering in which a
re-clustering is performed with the radius of  $R_{filt}=min(0.3,\Delta
R_{J_{1},\, J_{2}}/2)$ which selects at most three hardest jets. This is a
useful step to remove the contributions from pileup and underlying
events \cite{fatjetpileup}.
In the analysis, the two hardest objects are required to be tagged as b-jets
while the third one can be possible radiation of the two b quarks. 

It has been shown the best values
for the algorithm parameter are $\mu_{MD}=0.67$ and $y_{cut}=0.09$
\cite{fatjetSalam} and the best performance for clustering when it
deals with jet substructure is C/A algorithm.  It is worth mentioning
here that tighter value of $\mu_{MD}$  would not be useful significantly \cite{fatjetmasdrop}.
The algorithm explained above for reconstruction of the boosted objects
has been implemented in the FastJet3.1.1 package
\cite{fastjet} using that we perform the analysis to find the 
two Higgs bosons in the final state. 

In this analysis, first the jets coming from signal or backgrounds are
reconstructed with the anti-$k_{T}$ algorithm with the cone size of
$R=0.5$ then if two jets with $P_{T}>250$ GeV are found in the
event, the fat jet algorithm is applied. Otherwise, the event is
treated as normal event.

As for the b-tagging efficiency and mis-tag rates, similar numbers
as the CMS experiment are used. The data driven efficiency for the 
b-jet identification indicates that the efficiency is as large as
$60\%$ to  $80\%$ \cite{btag}.  In our analysis,  a b-tagging efficiency of
$70\%$ for jets with transverse momentum larger than 30 GeV and in the
pseudorapidity range of $|\eta| < 2.5$ is assumed. Mis-tagging rates
of $10\%$ and $1\%$ for the
c-jets and for the light jets are considered \cite{btag}.  

\subsection{Higgs bosons reconstruction\label{sec:Higgs-Mass-Reconstruction}}

\begin{figure}
\centering{}
\includegraphics[scale=0.35]{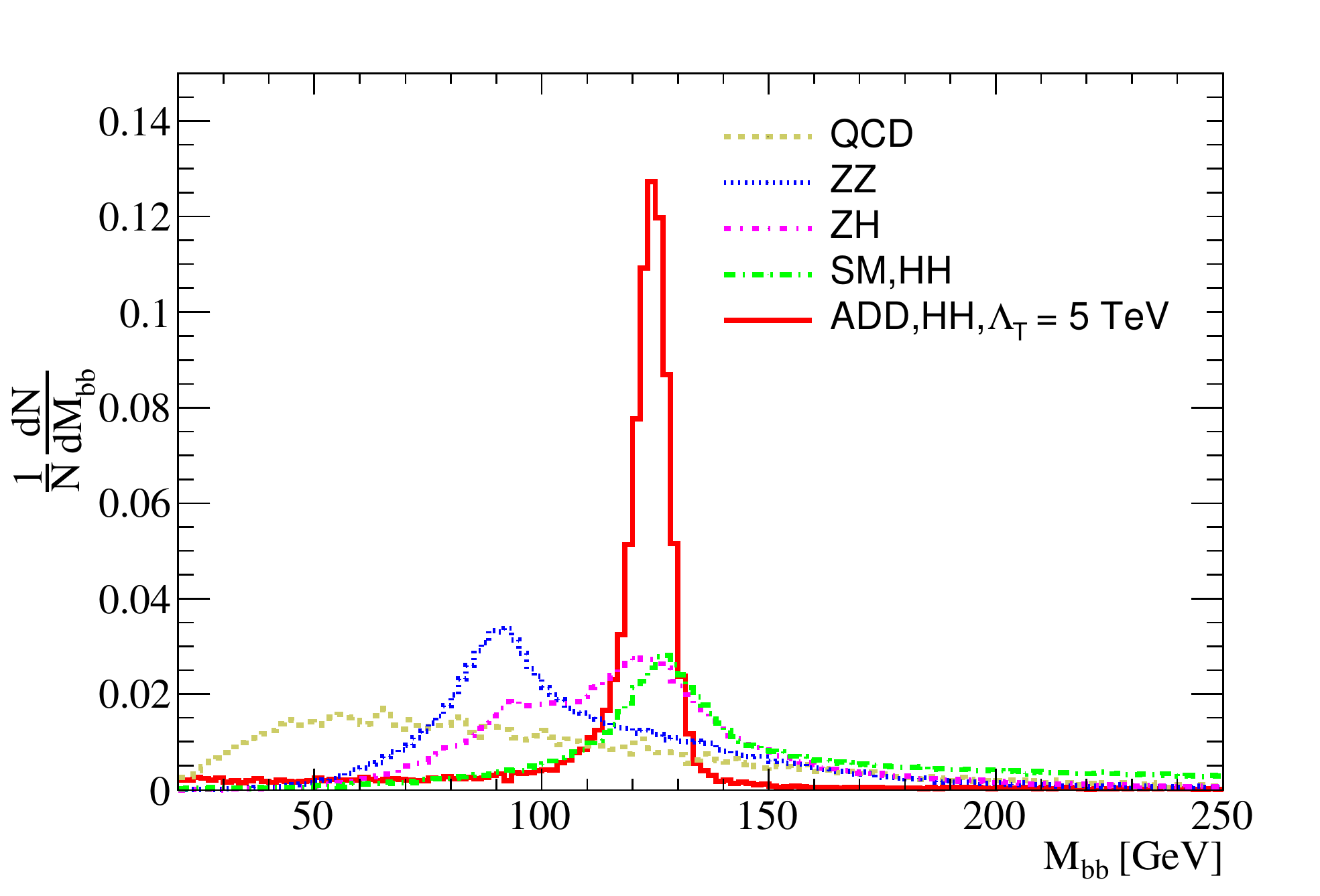}
\includegraphics[scale=0.35]{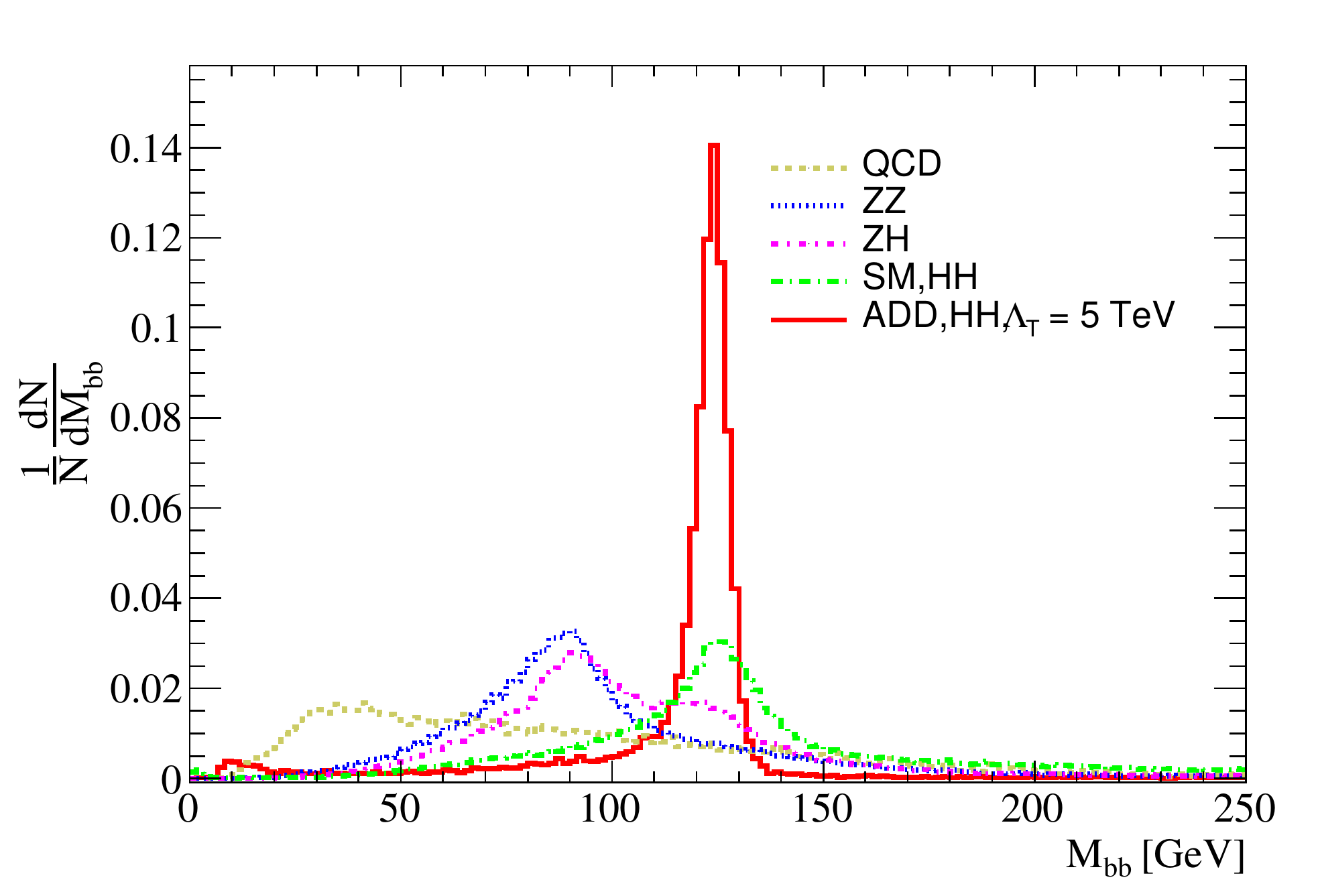}
\protect\caption{Normalized distributions of the reconstructed Higgs bosons for the signal
and backgrounds after applying the fat jet algorithm and detector
effects.\label{fig:reconstructedhiggs14}}
\end{figure}

For reconstruction of the Higgs bosons in the final state, 
a $\chi^{2}$ algorithm is utilized to determine the correct assignment
of b-jets to Higgs bosons candidates.  It relies on the Higgs boson
mass and other kinematics properties as constraints.  All possible permutations for four or more
b-jets are tried and the permutation with minimum $\chi^{2}$
is used to reconstruct both Higgs bosons. The $\chi^{2}$ which is run
over the events containing at least four b-jets with
$P_{T_{b}}>30$ GeV and $|\eta|<2.5$ is defined as:
\begin{equation}
\chi^{2}=(M_{ij}-M_{H})^{2}+(M_{kl}-M_{H})^{2}+\Delta R_{ij}^{2}+\Delta R_{kl}^{2}\label{eq:chisquarehiggs}
\end{equation}
where $M_{ij},\: M_{kl}$ are the invariant mass of the b-jets pairs
and $M_{H}=125$ GeV. As mentioned above, the best combination of b-jet pairs is the one
with minimum $\chi^{2}$ out of three possible combinations.
Figure \ref{fig:reconstructedhiggs14} shows the invariant mass of
the b-jet pairs after applying the fat jet algorithm and a CMS-like detector
effects. As it can be seen, the reconstructed Higgs bosons from signal have a
very good resolution on the mass spectrum.  For the sake of
comparison, the reconstructed mass distributions of  some
background processes are also shown in
Fig.\ref{fig:reconstructedhiggs14}.
Because of the efficient performance of the fat jet algorithm which
leads to a better resolution on the Higgs boson mass spectrum with
respect to all backgrounds, imposing cut on the invariant mass of each b-jet pairs can
suppress significant amount of the backgrounds keeping signal events
at a maximum level.

\begin{figure}
\begin{centering}
\includegraphics[scale=0.6]{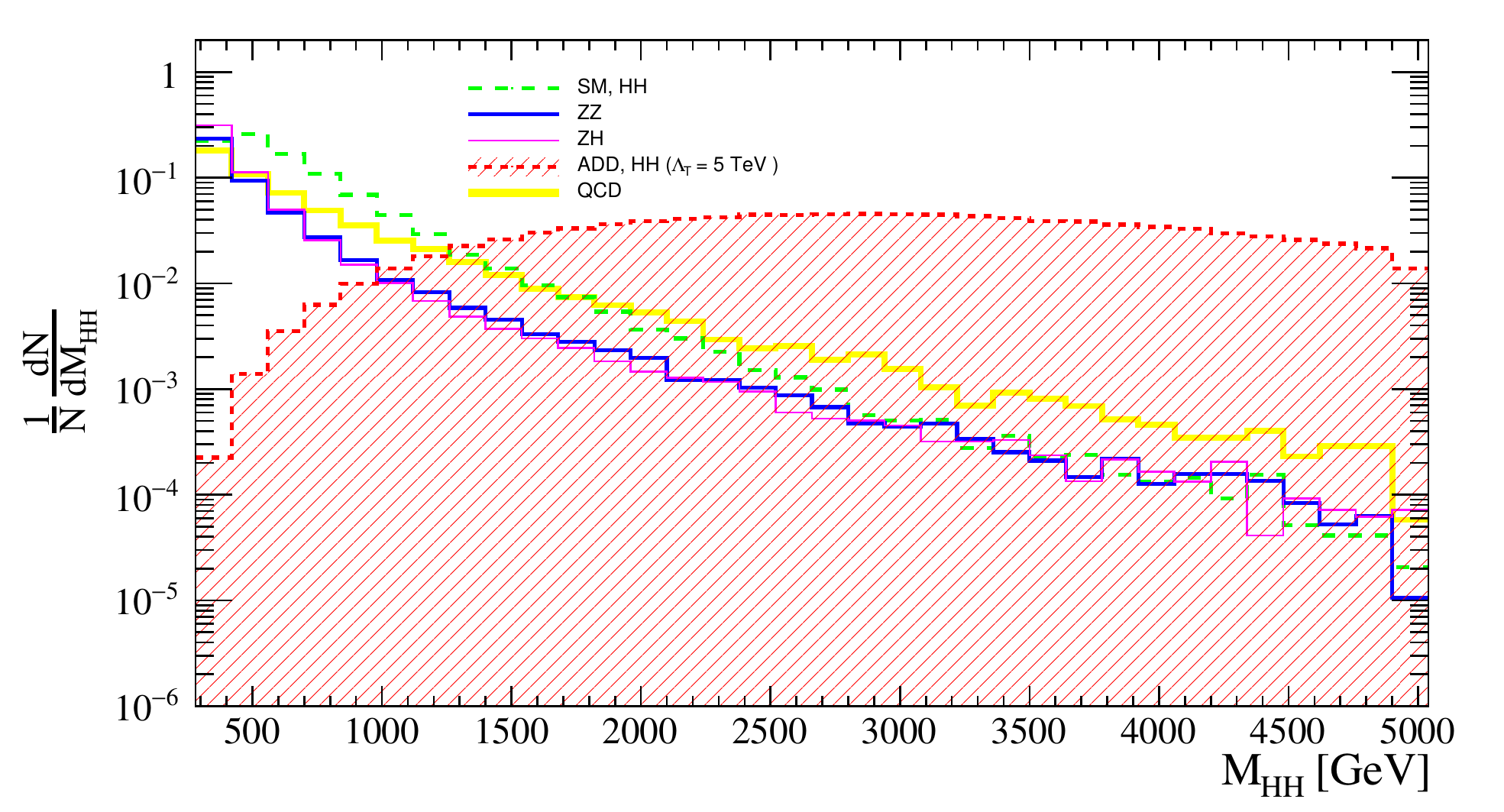}\protect\caption{Invariant mass of the Higgs pair for the signal sample with the parameter
of $\varLambda_{T}=5\, TeV$ and other sources of the backgrounds.
distribution is normalized to the number of the each samples separately.\label{fig:invmass14}}
\par\end{centering}
\end{figure}

According to discussions in the first section, we expect a continuous
enhancement in the rate of the signal events due to contribution
of the very close $G_{kk}$ modes of  gravitons.  This effect
manifests itself mostly in the high invariant mass region of di-Higgs
events where the number of excited modes of the $G_{kk}$ are much
larger.  This effect has been shown previously in Fig.\ref{fig:invdeltar},
the ADD signal events have very large di-Higgs invariant mass while
the SM di-Higgs events are distributed at lower invariant mass with
respect to signal. Such a discriminating feature is used in the next
section to set limits on the ADD signal model parameters.
Figure \ref{fig:invmass14} depicts the invariant mass of the two
reconstructed Higgs bosons for the signal and different sources of
backgrounds. This plot shows the behavior of the signal and potential
backgrounds in the final state mass spectrum. 
In this plot,  no cut except for the acceptance cuts are applied.

\subsection{Event selection\label{sec:Event-Selection}}

To select the  signal events, we require  events to have exactly
four b-tagged jets with $P_{T}>30$ GeV and $|\eta|<2.5$ and no
isolated lepton with $P_{T}>10$ GeV. The missing transverse energy is
required to be less than 10 GeV. All these requirements are denoted as cut-1.  
Additional cuts are applied for further backgrounds suppression.
Due to good resolution on angular separation between b-jet pairs
coming from each Higgs boson, we require each b-jets pair to have
$\Delta R<0.6$ (cut-2). Such a tight cut reduces the contributions of
non-boosted background events.  A mass window cut of 100 GeV to 150 GeV is applied
on the invariant mass of each reconstructed b-jets pair (cut-3) which
suppresses the contribution of the backgrounds with no Higgs boson.
We should mention here that triggering of the events with purely
hadronic final states needs careful attention however requiring four
energetic jets in the event has good enough efficiency at the LHC experiments.

Finally, we present the event yield at the center-of-mass energy of 14
TeV with 300 fb$^{-1}$ of the integrated luminosity in Table
\ref{tab:cutflowtable14tev-1}. In this table, the number of remaining
events are presented in a signal region of $M_{HH} > 1.4$
TeV for signal and main backgrounds. 
Table \ref{tab:cutflowtable14tev-1} explicitly shows
that in the high invariant mass of di-Higgs, i.e. the signal region,
the contribution of the reducible backgrounds are almost
negligible and the only main source of the background comes
from the SM Higgs pair production. As it can be seen and shown in
\cite{hhh2}, the combination of jet substructure techniques,
b-tagging requirement, and invariant mass cuts reduces the contribution of the QCD multijet
background negligible.   

The cut on the di-Higgs invariant
mass needs to be optimized to achieve best limits on the ADD model
parameter which will be explained in the next section.

\begin{table}
\begin{centering}
\begin{tabular}{|c|ccccccc|}
\hline 
 & \multicolumn{1}{c|}{{\scriptsize{}ADD,$\Lambda_{T}=6$ TeV}}& \multicolumn{1}{c|}{{\scriptsize{}SM,HH}}
 &\multicolumn{1}{c|}{{\scriptsize{}QCD}} &\multicolumn{1}{c|}{{\scriptsize{}Z$b\bar{b}$}} &
 \multicolumn{1}{c|}{{\scriptsize{}ZZ+WW}} & \multicolumn{1}{c|}{{\scriptsize{}ZH}} & {\scriptsize{}$t\bar{t}$}
\tabularnewline
\hline 
\hline 
{\scriptsize{} Cut on $M_{HH}$ } &  & & &
{\scriptsize{}$M_{HH} >$ 1.4 TeV} 
& & &
\tabularnewline
\hline 
\hline 
{\scriptsize{} cut-1} & {\scriptsize{}3.86} & {\scriptsize{}13.46} & {\scriptsize{}5.6e+5} & {\scriptsize{}1.2e+5} &
{\scriptsize{}26.47} & {\scriptsize{}4.34} & {\scriptsize{}40.70} \tabularnewline
\hline 
\hline 
{\scriptsize{} cut-2} & {\scriptsize{}3.51}& {\scriptsize{}1.96} & {\scriptsize{}13.48} & {\scriptsize{}31.0}&
{\scriptsize{}0.01} & {\scriptsize{}0.22} &  {\scriptsize{}0.00}\tabularnewline
\hline 
\hline 
{\scriptsize{} cut-3} & {\scriptsize{}2.28}& {\scriptsize{}0.55} & {\scriptsize{}0.02} & {\scriptsize{}0.30} &
{\scriptsize{}0.00} & {\scriptsize{}0.00} &  {\scriptsize{}0.00} \tabularnewline
\hline 
\end{tabular}
\par\end{centering}

\protect\caption{Number of survived events of ADD signal and different backgrounds
  after applying sets of cut-1, cut-2, and cut-3 described in the text
  in the signal region of $M_{HH}>1.4$ TeV
  for the  $\sqrt{s} = 14$ TeV and with 300 $fb{}^{-1}$ of integrated
  luminosity of expected data from the
LHC.\label{tab:cutflowtable14tev-1}}
\end{table}

\section{Limit calculation}

In this section we present the statistical procedure that we use to
obtain the expected limits on the ADD model parameter. As we mentioned
formerly,  the invariant mass of the Higgs pair is an effective
observable that we use to set  limit on the signal cross section and
then translate the limit on the model parameters in the absence of any
indication of the ADD model
signal.  A single bin counting experiment in the signal
dominant region (high invariant mass region) is used to set the
limits. We begin with a Poisson distribution as the probability of
measuring $N$ events in the signal region:

\begin{equation}
P(N|\,\sigma_{ADD}\,\varepsilon\,\mathcal{L} ,
B)=e^{-(B+\sigma_{ADD}\varepsilon\mathcal{L} )}\frac{(B+\sigma_{ADD}\varepsilon\mathcal{L})^{N}}{N!}\label{eq:likelohood}
\end{equation}
where $\sigma_{ADD},\:\varepsilon,\:\mathcal{L}\:\ $ and $B$ are signal cross section,
signal efficiency, integrated luminosity and expected number of backgrounds.
In the above equation,  $\sigma_{ADD}$ is taken as a free parameter to
enable us to consider different ADD signal production cross sections.
 To obtain the number of expected background events $B$ and the signal
 efficiency $\epsilon$, we rely on the Monte Carlo simulations.
At confidence level of $95\%$, an upper limit on the signal
rate $\sigma_{ADD}$ is obtained by integrating over the
posterior probability as the following:
\begin{equation}
0.95=\frac{\int_{0}^{\sigma^{95\%}}P(N|\,\sigma_{ADD}\,\varepsilon\,\mathcal{L},
  B)d\sigma_{ADD}}{\int_{0}^{\infty}P(N|\,\sigma_{ADD}\,\varepsilon\,\mathcal{L},  B) d\sigma_{ADD}}\label{set}
\end{equation}
In order to extract the expected limit on the ADD signal cross
section, one has to solve the Eq.\ref{set} for $\sigma^{95\%}$
under the assumption of $N=B$ after inserting the proper inputs
for the background expectation, signal efficiency, and the integrated luminosity.

\begin{figure}
\begin{centering}
\includegraphics[scale=0.46]{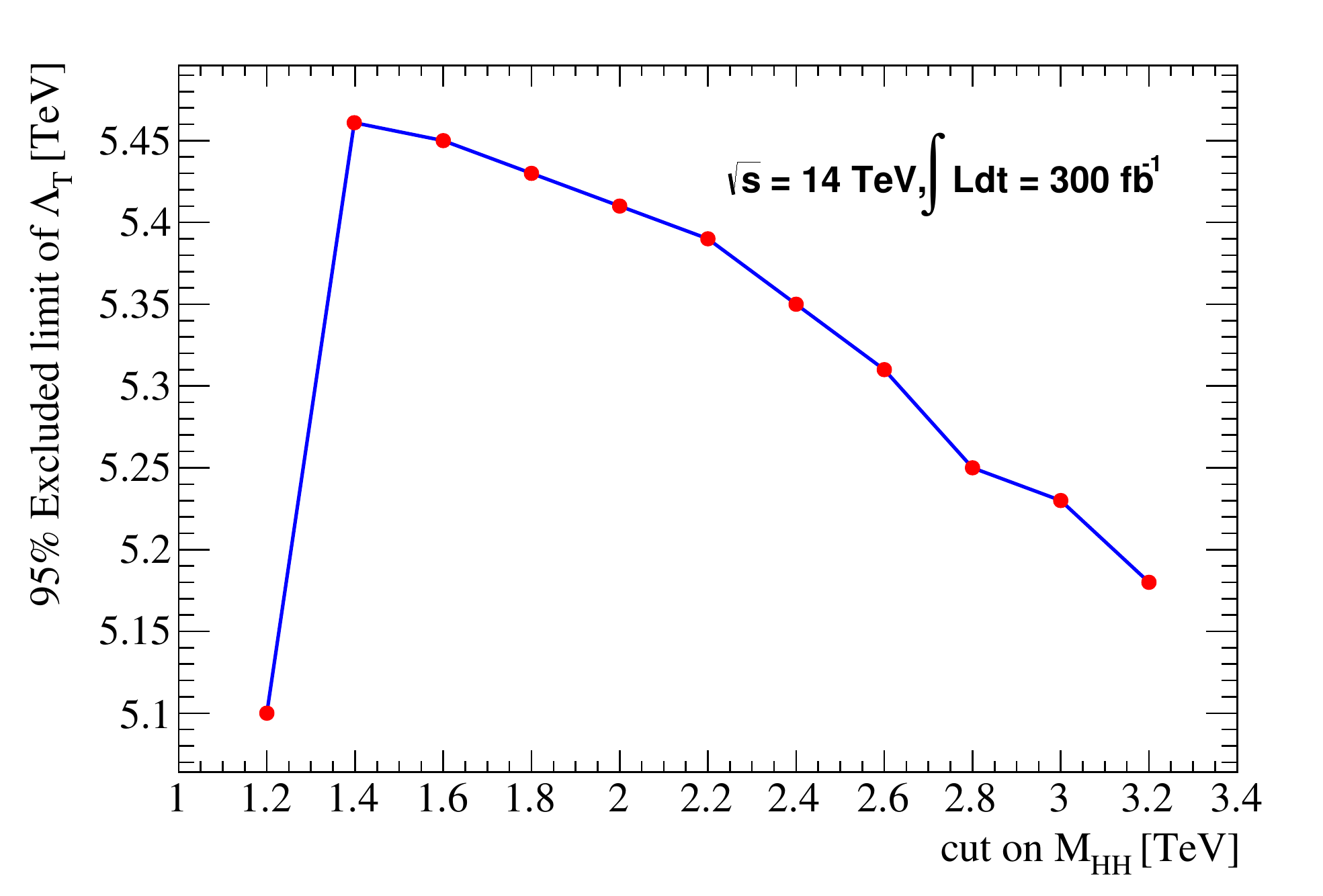}
\par\end{centering}

\centering{}\protect\caption{The $95\%$ CL expected limit on $\varLambda_{T}$ as a function
of cut on the invariant mass of two reconstructed  Higgs bosons.\label{fig:optimummass}}
\end{figure}

In the first step of limit setting,   we have to determine the signal
dominant region. Therefore, the di-Higgs invariant mass cut which
determines this region is optimized in such a way
that gives the best exclusion limits on the model parameter
$\varLambda_{T}$.  This can be reached by minimizing the $95\%$ CL expected limit on the signal
cross section. Figure \ref{fig:optimummass} shows the calculated
 expected limit at $95\%$ CL on $\varLambda_{T}$ as a function of invariant
mass cut of di-Higgs. As it can be seen in Fig.\ref{fig:optimummass},
with increasing the cut on di-Higgs mass the exclusion limit on
$\Lambda_{T}$ is maximized at the cut on di-Higgs mass of around 1.4
TeV with an integrated luminosity of 300 fb$^{-1}$.
Thus,  we take the mass cut of 1.4 TeV as the optimized value to
introduce the signal region. It has to be mentioned that
in the optimization process no systematic uncertainty is included. It
is worth mentioning that the optimized cut on the di-Higgs invariant
mass varies with the integrated luminosity.

We calculated the signal efficiency after applying the set of cuts
described in the previous section. This efficiency has almost a flat
behavior against the model parameter $\Lambda_{T}$ . The mean value
of the signal efficiency is taken to calculate the limit. It is found
to be equal to $17\%$. The uncertainty on the efficiency originating
from statistical fluctuations and a $5\%$ uncertainty due to the fluctuation of efficiency
for different $\Lambda_{T}$ are considered.
An overall uncertainty of $30\% $ on the number of background events
in addition to the statistical uncertainty is considered.

\begin{figure}
\centering{}\includegraphics[scale=0.46]{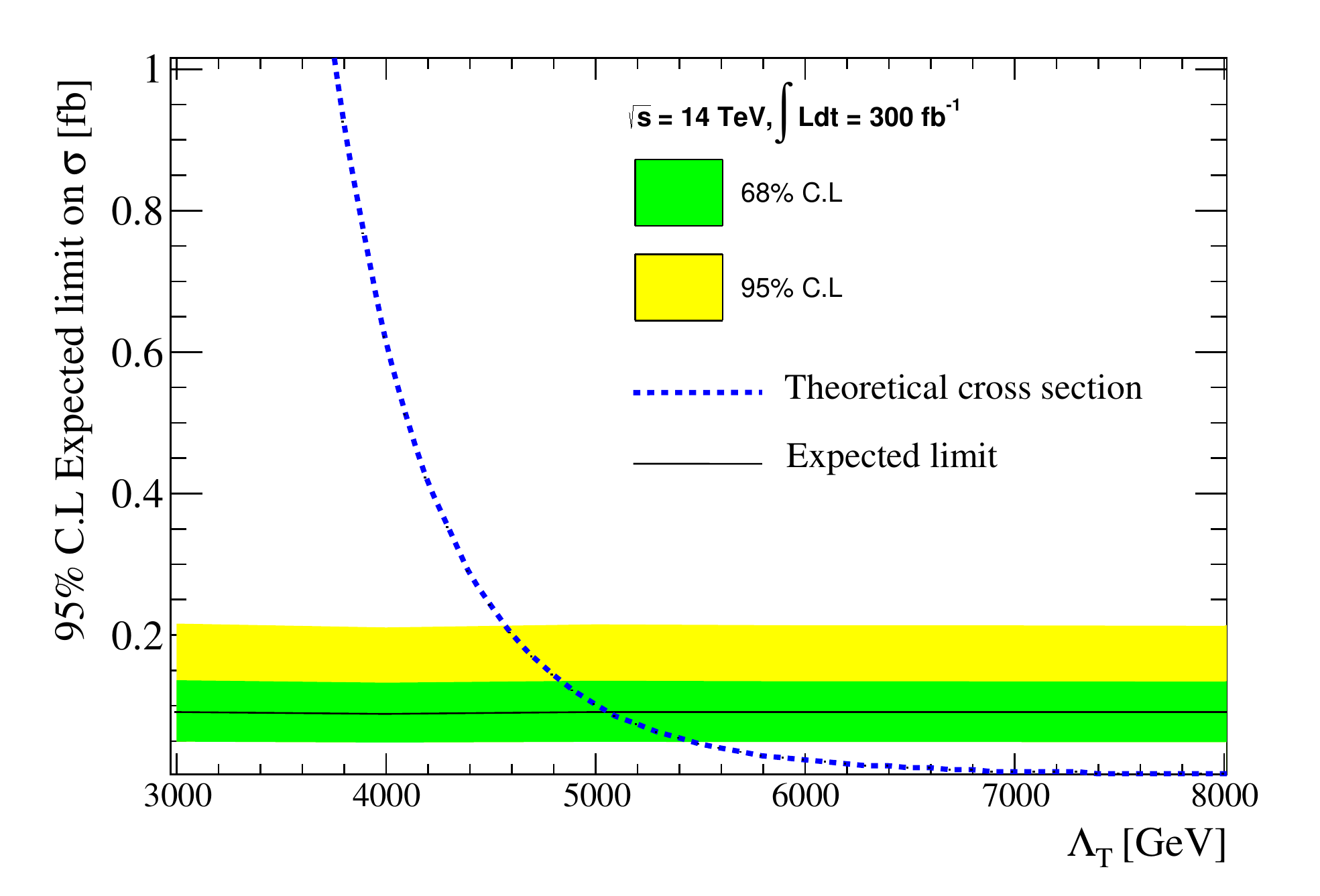}\protect\caption{The
  $95\%$ CL expected limit on the ADD signal cross section as a function of $\Lambda_{T}$
including the uncertainty bands. The theoretical cross section of ADD model
is also presented for comparison.\label{fig:limit14tev300fb}}
\end{figure}

Figure \ref{fig:limit14tev300fb} shows the expected
limit at $95\%$ CL as a function of model parameter $\Lambda_{T}$
including the uncertainty bands.
The theoretical cross section of ADD signal also is presented for comparison.
The $95\%$ CL expected upper limit on signal cross section in the signal
region is found to be $0.09$ fb for an integrated luminosity of 300
fb$^{-1}$ of data. It leads to an expected lower limit on
$\Lambda_{T}$ to be 5.1 TeV. These bounds are in a reasonable
agreement with the $3\sigma$ exclusion limits calculated in \cite{hh1}
where no object reconstruction and identification and detector effects have
been considered.

Similar analyses are performed for higher center-of-mass energies of
future planed proton-proton colliders with the integrated
luminosities of 300 fb$^{-1}$ and 3 ab$^{-1}$. The results for expected limits on the $\Lambda_{T}$ in GRW convention
and on ($n{}_{ED}$, $M{}_{S}$) in the HLZ convention are summarized in
Tables \ref{tab:300fblimittable} and \ref{tab:3ab}, respectively.
Moving to larger center-of-mass energy of the collisions leads to
increase the lower limit on $\Lambda_{T}$. The limit is extended up to
around 24 TeV at the collision energy of 100 TeV. Using more
integrated luminosity of data would lead to improve the limit on the
model parameter. Increasing the integrated luminosity by a factor of
10 changes the lower limit on $\Lambda_{T}$ from 5.1 TeV to 6.8 TeV at
$\sqrt{s} = 14$ TeV.

At the end of this section, it must be mentioned that including the
other decay channels of the Higgs bosons
($\gamma\gamma,\tau^{+}\tau^{-},ZZ,WW$) would improve the limits
considerably. 

\begin{table}
\begin{centering}
\begin{tabular}{|c|c|c|}
\hline 
$\begin{array}{c}
IL=300\; fb^{-1}\\
\text{center-of-mass~energy}
\end{array}$ & %
\begin{tabular}{c}
$\varLambda_{T}\:[TeV] (GRW)$\tabularnewline
\end{tabular} & \textcompwordmark{}$\begin{array}{c}
M_{S}\:[TeV] (HLZ)\\
n=3\; n=4\; n=5\; n=6\; n=7
\end{array}\:$\tabularnewline
\hline 
$14$ TeV & 5.1 & $6.1\qquad5.1\qquad4.6\qquad4.3\qquad4.1$\tabularnewline
\hline 
$33$ TeV & 10.5 & $12.5\quad\;\;10.5\quad\;\;9.5\quad\;\;8.8\quad\;\;8.3$\tabularnewline
\hline 
$100$ TeV & 23.6 & $28.1\quad\;\;23.6\;\quad\;21.3\;\quad\;19.8\;\quad\;18.8$\tabularnewline
\hline 
\end{tabular}\protect\caption{$95\%$ CL expected limit on the parameters of the model in both GRW
and HLZ conventions for the 300 fb $^{-1}$ of integrated luminosity of
data. \label{tab:300fblimittable}}
\par\end{centering}
\end{table}

\begin{table}
\centering{}%
\begin{tabular}{|c|c|c|}
\hline 
$\begin{array}{c}
IL=3\; ab^{-1}\\
\text{center-of-mass~energy}
\end{array}$ & %
\begin{tabular}{c}
$\varLambda_{T}\:[TeV] (GRW)$\tabularnewline
\end{tabular} & \textcompwordmark{}$\begin{array}{c}
M_{S} \:[TeV] (HLZ)\\
n=3\; n=4\; n=5\; n=6\; n=7
\end{array}\:$\tabularnewline
\hline 
$14$ TeV & 6.8 & $8.1\qquad6.8\qquad6.1\qquad5.7\qquad5.4$\tabularnewline
\hline 
$33$ TeV & 13.4 & $16.0\quad\;\;13.4\quad\;\;12.1\quad\;\;11.3\quad\;\;10.7$\tabularnewline
\hline 
$100$ TeV & 28.7 & $34.1\quad\;\;28.7\;\quad\;25.9\;\quad\;24.1\;\quad\;22.8$\tabularnewline
\hline 
\end{tabular}\protect\caption{$95\%$ CL expected limit on the
  parameters of the ADD model in both GRW
and HLZ conventions for the 3 ab$^{-1}$data.\label{tab:3ab}}
\end{table}

\section{Di-Higgs angular distribution}

An interesting feature of di-Higgs production from the large extra
dimensions is the quite different behavior of the angular distribution
of the final state with respect to the SM backgrounds.
In general,  final state particles coming from the exchange of gravitons
with spin 2  should have different shape from the final state
particles from the exchanges of  photon, $Z$-boson or Higgs boson. On
the other hand,  using fat jet algorithm enabled us to have very good
resolution on angular separation. Therefore, angular distribution of
the Higgs boson pairs can be used as a powerful observable to distinct the
ADD signal from the SM backgrounds to set limits on the model parameters.

\begin{figure}
\begin{centering}
\includegraphics[scale=0.46]{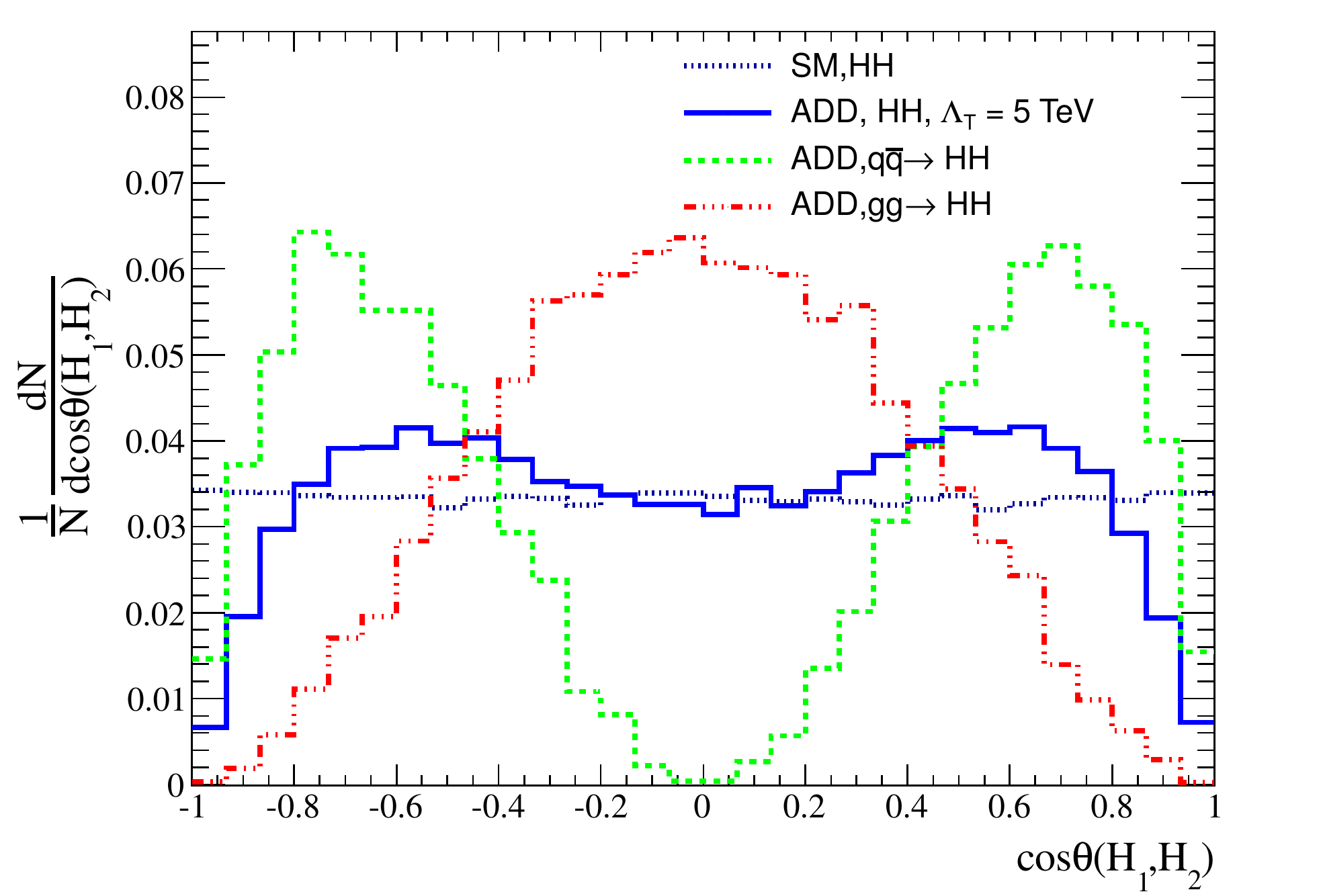}\protect\caption{The
  distribution of $\cos\theta(H_{1},H_{2})$ for the SM di-Higgs background and
  the extra dimension signal at parton level.\label{fig:costeta}}
\par\end{centering}
\end{figure}

The shape of the angular distribution of di-Higgs, which is 
an interesting feature of the ADD model, is shown in
Fig.\ref{fig:costeta}.
The angular distribution of the SM di-Higgs is
presented for comparison.
In this plot, $\theta(H_{1},H_{2})$ is the angle between the directions of
the momenta of the final state Higgs bosons. The distribution of
$\cos\theta(H_{1},H_{2})$ is plotted for ADD di-Higgs production from $q\bar{q}$
annihilation and $gg$ fusion separately. As it can be seen, the angular
distribution of the signal events from $gg$ fusion has quite different
behavior from $q\bar{q}$. The two Higgs of ADD events produced from
$gg$ prefer to fly mostly perpendicular to each other while the two
Higgs bosons in the ADD events from $q\bar{q}$ fusion tend to be produced at the
angles of approximately  $ \pm \pi/4$. Detailed analytical
explanations of the angular distributions of di-Higgs from the ADD
model in $e^{+}e^{-}$ collisions and $\gamma\gamma$ collisions can be
found in \cite{eehh} and \cite{gagahh}.  Similar explanations are valid
for the hadron colliders with the initial states of  $q\bar{q}$ and
$gg$ are expected to be like $e^{+}e^{-}$  and $\gamma\gamma$, respectively.
According to Fig.\ref{fig:costeta} , the SM di-Higgs distribution is quite flat and have different shape
from the ADD signal events. 
It is worth mentioning here that as discussed in section 1,  due to
larger gluon PDF at larger center-of-mass energies the contribution of
$gg$ fusion in ADD signal production is increased. It has been shown
in Fig. \ref{fig:xsecgg}. 

The CMS and ATLAS collaborations \cite{rp1,rp2} have used a variable of $X=exp(|y_{1}-y_{2}|)$
to search for the contact interactions in di-jet events. The
rapidities of the two jets are denoted by $y_{1},y_{2}$.  The rapidity
$y$ is defined as $\log(\frac{E+p_{z}}{E-p_{z}})$ where $E$ is the
energy and $p_{z}$ is the $z$-component of the momentum of a given
particle. The advantage of the rapidity difference is that it is a
boost invariant quantity. Now, we use the
$X=exp(|y_{H_{1}}-y_{H_{2}}|)$ distribution to probe the effects of
ADD model instead of $\cos\theta(H_{1},H_{2})$.

\begin{figure}
\begin{centering}
\includegraphics[scale=0.46]{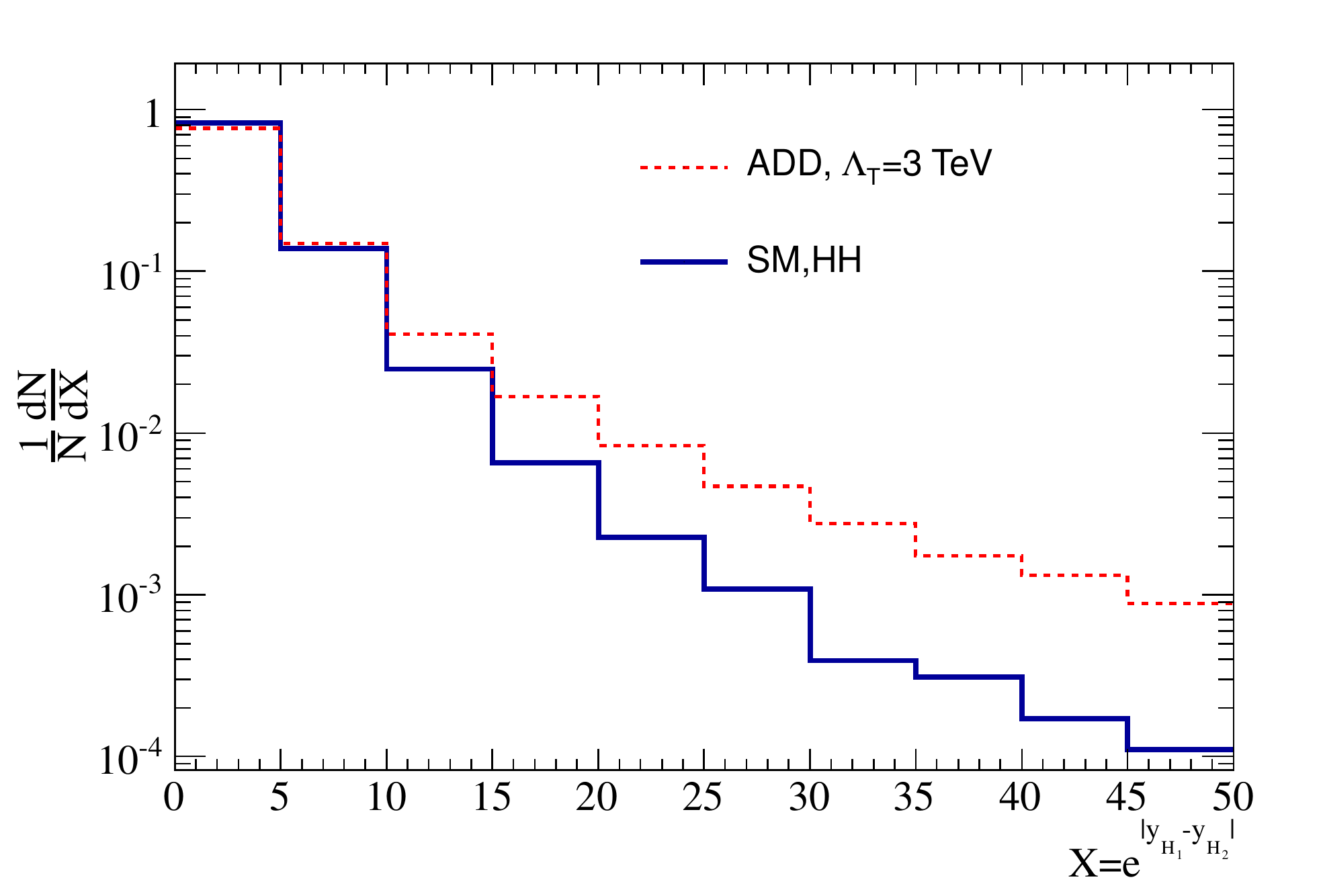}\protect\caption{ The
  normalized distribution of $\exp(y_{H_{1}}-y_{H_{2}})$ for the ADD
  signal events and SM di-Higgs production at parton level.\label{fig:rapiditygap}}
\par\end{centering}
\end{figure}

\begin{figure}
\begin{centering}
\includegraphics[scale=0.46]{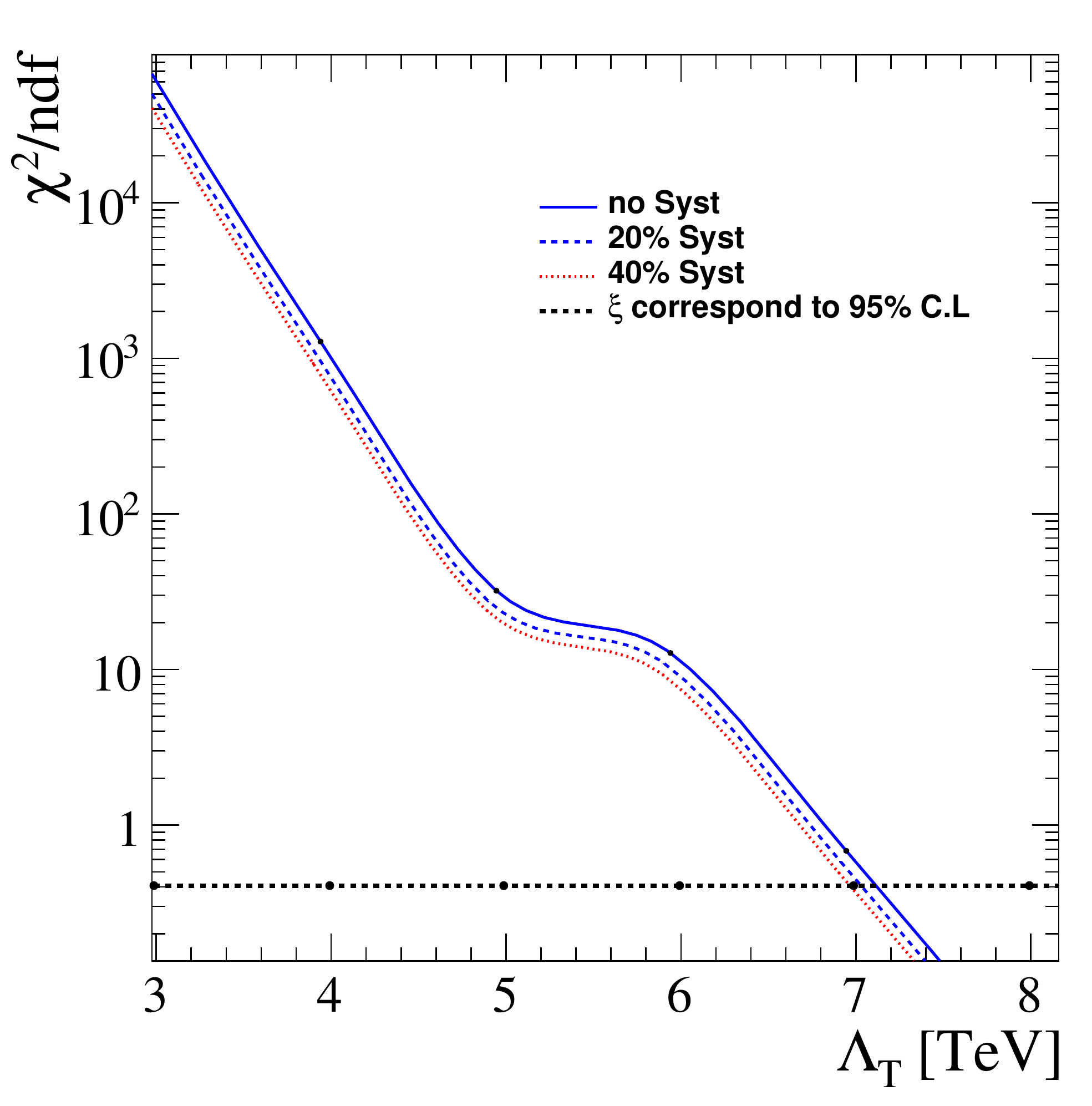}\protect\caption{The $\chi^{2}/n$ as a function of $\Lambda_{T}$, where $n$ represents
the number of degrees of freedom. The dashed line shows the
value of $\chi^{2}/n$ which corresponds to the $95\%$ CL.\label{fig:chisquare}}
\par\end{centering}
\end{figure}

Figure \ref{fig:rapiditygap} shows of $X=exp(|y_{H_{1}}-y_{H_{2}}|)$
for main SM background di-Higgs and the ADD signal with $\Lambda_{T} =
3$ TeV.  Considering only the SM di-Higgs as the main background, we
set a limit on the ADD parameter using this angular distribution. We
perform the analysis with an integrated luminosity of 300 fb$^{-1}$. 
We define a $\chi^{2}$ over the $X$ distribution as:

\begin{equation}
\chi^{2}=\sum_{i=0}^{n}\frac{(N_{sig+bck}-N_{bck})^{2}}{\sqrt{\Delta_{stat}^{2}+\Delta_{syst}^{2}}}
\end{equation}
where the numerator shows the square number of signals and $\Delta_{stat}$
and $\Delta_{syst}$ represent the statistical and systematical uncertainties. 
To calculate the $\chi^{2}$, we use the same event selections
as before and we consider the events in the signal region which is
determined in the previous section. Figure \ref{fig:chisquare} shows
the $\chi^{2}/n$ as a function of $\Lambda_{T}$, where $n$ indicates
the number of degree of freedom. The dashed-line shows the
value of $\chi^{2}/n$ which corresponds to the $95\%$ CL. 
The limit on $\Lambda_{T}$ using this observable is found to
be 6.98 TeV which is higher than the value that we obtained
using the invariant mass. To check the effect of 
systematic uncertainties, we considered $20\%$ and $40\%$ overall systematic
uncertainties which are shown as dashed lines. Including $40\%$
systematic uncertainty leads to loosen the limit on $\Lambda_{T}$
around 200 GeV. A significant improvement could be achieved using the
distribution of the rapidity gap of two-Higgs bosons which amounts to
around 1.8 TeV with respect to the mass spectrum analysis of
di-Higgs. 

This leads to conclude that the di-Higgs final state
would be a promising channel to search for the large extra dimension effects
at the hadron colliders. In particular, the usage of angular
distribution would lead to stringent bounds on the model parameters.

\section{Summary and conclusions}

Double Higgs boson production at hadron colliders provides the possibility to
probe not only the Higgs self-coupling and Higgs couplings with the SM
particles but also it enables us to search for the effects of new
physics beyond the SM. In this paper, the double Higgs production at
the LHC and future circular collider (FCC) with the center-of-mass
energies of 14 TeV, 33 TeV, and 100 TeV is used to search for the
effects of large extra dimensions. The analysis is only based on the
most probable final state i.e. $pp \rightarrow HH \rightarrow
b\bar{b}b\bar{b}$ which is of course a challenging channel due to
large QCD background and triggering the events. The tail of the invariant mass of
the two Higgs bosons is affected due to the virtual gravitons
exchange.  In addition to the di-Higgs invariant mass, the angular
distributions of the final state Higgs bosons (and consequently the
decay products) have a quite different behavior with respect to the
SM irreducible background due to the exchange of
spin 2 gravitons.  We perform a comprehensive Monte-Carlo simulation
analysis taking into account the main backgrounds and consider a
CMS-like detector effects using the Delphes package. To reconstruct
the signal candidate events efficiently and reasonable background
rejection, jet substructure techniques are employed to capture
the signal events which are boosted objects in the final state. Then
we obtain the expected limits on the model parameters using the invariant
mass and the angular properties of the final state
particles. Depending on the number of extra dimensions,  the effective
Planck scale is limited up to  6.1, 12.5, 28.1 TeV at the proton-proton collisions
with the center-of-mass energies of 14, 33, and 100 TeV, respectively. 
Further improvement of the analysis is possible by including other
decay modes of the Higgs bosons such as
$\gamma\gamma,WW,ZZ,\tau^{+}\tau^{-}$.

{\bf Acknowledgement}
The authors would like to thank Sherpa  authors for their technical
helps in ADD signal event generations.


\end{document}